\documentstyle[12pt,epsf]{article}

\begin{document}

\baselineskip 6mm
\renewcommand{\thefootnote}{\fnsymbol{footnote}}

%------------ Hyun Seok's macro's, etc  -----------

\newcommand{\nc}{\newcommand}
\newcommand{\rnc}{\renewcommand}

%\headheight=0truein
%\headsep=0truein
%\topmargin=0truein
%\oddsidemargin=0truein
%\evensidemargin=0truein
%\textheight=9truein
%\textwidth=6.5truein

\rnc{\baselinestretch}{1.24}    % 1.5 spacing btwn text lines
\setlength{\jot}{6pt}       % spacing btwn the rows of an eqnarray
\rnc{\arraystretch}{1.24}   % spacing btwn the rows of a non-eqn array

%%%%%%%%%%%%%%%%%%%%%% Equation Numbering %%%%%%%%%%%%%%%%%%%%%%%
\makeatletter
\rnc{\theequation}{\thesection.\arabic{equation}}
\@addtoreset{equation}{section}
\makeatother

%%%%%%%%%%%%%%%%%%%%%%%%%%%%%%%%%%%%%%%%%%%%%%%%%%%%%%%%%%%%%%%%%
%                                                               %
%                NEW COMMANDS AND MACROS                        %
%                                                               %
%%%%%%%%%%%%%%%%%%%%%%%%%%%%%%%%%%%%%%%%%%%%%%%%%%%%%%%%%%%%%%%%%

%%%%% Simplify some frequently used LaTeX commands %%%%%

\nc{\be}{\begin{equation}}

\nc{\ee}{\end{equation}}

\nc{\bea}{\begin{eqnarray}}

\nc{\eea}{\end{eqnarray}}

\nc{\xx}{\nonumber\\}

\nc{\ct}{\cite}

\nc{\la}{\label}

\nc{\eq}[1]{(\ref{#1})}

\nc{\newcaption}[1]{\centerline{\parbox{6in}{\caption{#1}}}}

\nc{\fig}[3]{

\begin{figure}
\centerline{\epsfxsize=#1\epsfbox{#2.eps}}
\newcaption{#3. \label{#2}}
\end{figure}
}

%%% Caligraphic letters %%%%

\def\CA{{\cal A}}
\def\CC{{\cal C}}
\def\CD{{\cal D}}
\def\CE{{\cal E}}
\def\CF{{\cal F}}
\def\CG{{\cal G}}
\def\CH{{\cal H}}
\def\CK{{\cal K}}
\def\CL{{\cal L}}
\def\CM{{\cal M}}
\def\CN{{\cal N}}
\def\CO{{\cal O}}
\def\CP{{\cal P}}
\def\CR{{\cal R}}
\def\CS{{\cal S}}
\def\CU{{\cal U}}
\def\CW{{\cal W}}
\def\CY{{\cal Y}}

%%% Double line letters %%%

\def\IR{{\hbox{{\rm I}\kern-.2em\hbox{\rm R}}}}
\def\IB{{\hbox{{\rm I}\kern-.2em\hbox{\rm B}}}}
\def\IN{{\hbox{{\rm I}\kern-.2em\hbox{\rm N}}}}
\def\IC{\,\,{\hbox{{\rm I}\kern-.59em\hbox{\bf C}}}}
\def\IZ{{\hbox{{\rm Z}\kern-.4em\hbox{\rm Z}}}}
\def\IP{{\hbox{{\rm I}\kern-.2em\hbox{\rm P}}}}
\def\IH{{\hbox{{\rm I}\kern-.4em\hbox{\rm H}}}}
\def\ID{{\hbox{{\rm I}\kern-.2em\hbox{\rm D}}}}

%%% Greek letters %%%

\def\a{\alpha}
\def\b{\beta}
\def\ga{\gamma}
\def\d{\delta}
\def\ep{\epsilon}
\def\ph{\phi}
\def\k{\kappa}
\def\l{\lambda}
\def\m{\mu}
\def\n{\nu}
\def\th{\theta}
\def\rh{\rho}
\def\s{\sigma}
\def\t{\tau}
\def\w{\omega}
\def\G{\Gamma}

%%%%% Mathematical Symbols

\def\half{\frac{1}{2}}
\def\dint#1#2{\int\limits_{#1}^{#2}}
\def\goto{\rightarrow}
\def\para{\parallel}
\def\brac#1{\langle #1 \rangle}
\def\grad{\nabla}
\def\curl{\nabla\times}
\def\div{\nabla\cdot}
\def\p{\partial}
\def\e{\epsilon_0}

%%%%% Roman pont in math

\def\Tr{{\rm Tr}\,}
\def\det{{\rm det}}

%%%%% Special Letters

\def\vare{\varepsilon}
\def\barz{\bar{z}}
\def\barw{\bar{w}}

%%%%% For this paper only

\def\ad{\dot{a}}
\def\bd{\dot{b}}
\def\cd{\dot{c}}
\def\dd{\dot{d}}
\def\so{SO(4)}
\def\sop{SO(4)^\prime}
\def\bc{{\bf C}}
\def\bfz{{\bf Z}}
\def\bz{\bar{z}}

\begin{titlepage}

%---------------- preprint number ---------------

\hfill\parbox{5cm} {SOGANG-HEP 308/03 \\
SNUTP 03-014 \\
{\tt hep-th/0307146}}

\vspace{25mm}

\begin{center}
%------------------------ title ------------------------
{\Large \bf Intersecting D-branes in Type IIB Plane Wave Background}

\vspace{15mm}
%---------------- authors and addresses ----------------
Kyung-Seok Cha$^{\, a \,}$\footnote{quantum21@korea.com},
Bum-Hoon Lee$^{\, a \,}$\footnote{bhl@ccs.sogang.ac.kr}
and Hyun Seok Yang$^{\, b \,}$\footnote{hsyang@phya.snu.ac.kr}
\\[10mm]

${}^a${\sl Department of Physics, Sogang University,
Seoul 121-742, Korea} \\
${}^b${\sl School of Physics, Seoul National University,
Seoul 151-747, Korea} \\
\end{center}

\thispagestyle{empty}

\vskip2cm

%----------------------- abstract ----------------------

\centerline{\bf ABSTRACT}
\vskip 4mm
\noindent

We study intersecting D-branes in a type IIB plane wave background
using Green-Schwarz worldsheet formulation. We consider all
possible $D_\pm$-branes intersecting at angles in the plane wave background
and identify their residual supersymmetries. We find, in particular, that
$D_\mp - D_\pm$ brane intersections preserve no supersymmetry.
We also present the explicit worldsheet expressions of conserved
supercharges and their supersymmetry algebras.
\\


\vspace{1cm}

\today

\end{titlepage}

\renewcommand{\thefootnote}{\arabic{footnote}}
\setcounter{footnote}{0}

%%%%%%%%%%%%%%%%%%%%%%%%%%%%%%%%%%%%%%%%%%%%%%%%%%%%%%%%%%%%%%%%%%%%%%
\section{Introduction}
%%%%%%%%%%%%%%%%%%%%%%%%%%%%%%%%%%%%%%%%%%%%%%%%%%%%%%%%%%%%%%%%%%%%%%

The Penrose limit of the $AdS_5 \times S^5$ background in type
IIB supergravity corresponds to a plane wave solution \ct{blau},
\bea \la{pp-metric}
&& ds^2 = -2 dx^+ dx^- - \mu^2 x_I^2 (dx^+)^2 + {dx_I}^2, \\
&& F_{+1234}=F_{+5678}=2 \mu. \nonumber
\eea
This implies a correspondence between type IIB string theory in
the plane wave background \eq{pp-metric} and $\CN=4$
supersymmetric Yang-Mills theory with large R-charge. Since the
background \eq{pp-metric} is one of the very few Ramond-Ramond backgrounds on
which string theory is exactly solvable \ct{metsaev1,metsaev2},
one may have a genuine
hope to explicitly check the conjectured AdS/CFT correspondence
beyond the supergravity approximation on the string theory side.
Indeed Berenstein, Maldacena, and Nastase \ct{bmn} succeeded in
reproducing the string spectrum from perturbative super Yang-Mills
theory, thereby putting the duality on a firm ground at the free
theory level. Subsequent developments using the super Yang-Mills
theory and the light-cone string field theory have accumulated
strong evidences that the duality is still valid even after the
interactions both on the super Yang-Mills theory side and on the
string theory side are introduced.

D-branes can be described by boundary states of closed string
state. The symmetries that the boundary state preserves are
generically the combinations of the closed string symmetries that
leave the boundary state invariant. However, in the plane wave
background \eq{pp-metric}, it was shown by Skenderis and Taylor
in \ct{skenderis1,skenderis2} that an open string on a $D_+$-brane
has a different kind of kinematical supersymmetry not descending
from the closed string. Furthermore it was shown in \ct{hikida,ggsns}
that oblique D-branes exist in the background \eq{pp-metric} whose
isometry is a subgroup of the diagonal $SO(4)$ symmetry of the
background. Recently possible D-branes in plane wave backgrounds
have been identified
and their supersymmetries have also been classified \ct{skenderis1}-\ct{kimpark}.

The classification of D-branes in plane wave backgrounds is still
incomplete. In particular possible intersecting D-branes are not
completely studied even in the type IIB plane wave background
\eq{pp-metric}. $Dp-Dq$ brane systems have
an important role in string theory since they probe the
nonperturbative dynamics of the string theory and they have been
used to study various duality aspects of the string theory.
Furthermore intersecting D-branes have received intense interests
since it has been known that chiral fermions can appear on the
intersection of D-branes \ct{bdl} and so they are promising tools to
construct a phenomenological model like the Standard Model in
string theory context. Some aspects have been reviewed in
\ct{polchinski,smith}. Also recently there have been many works on
intersecting D-branes in other plane wave
and AdS backgrounds \ct{seki}-\ct{ohta}.

In this paper we consider all possible $D_\pm$-branes intersecting at angles
in the plane wave background \eq{pp-metric} and identify their residual
supersymmetries. It turns out that in some cases intersecting
D-branes are similar to the flat spacetime case but in other cases
they are quite different from the flat spacetime case. To
study the intersecting D-branes in the background \eq{pp-metric},
we develop the systematic method using the Green-Schwarz
superstring action in light-cone gauge for an open string
stretched between a $Dp$-brane and a $Dq$-brane. Our results
consistently recover
those in flat spacetime \ct{polchinski} when the limit $\mu \to 0$, namely,
the flat spacetime, is taken. Up to our best knowledge, our
methodology for studying intersecting D-branes, namely, the
worldsheet formalism using Green-Schwarz superstring action, is
new even in the flat spacetime case. Most relevant works may be \ct{gg,lambert}.
A merit of this formalism is that the spacetime supersymmetry of
intersecting D-branes is manifest.

This paper is organized as follows. In Sec. 2, we present a
worldsheet formulation using the Green-Schwarz superstring action
in light-cone gauge for an open string stretched between a
$Dp$-brane and a $Dq$-brane which are either parallel or
intersecting at right angles. We get the mode expansion of open
strings consistent with open string boundary conditions in the
Green-Schwarz superstring theory context. We find
that $D_\mp - D_\pm$ brane intersections preserve no supersymmetry.
In Sec. 3, the analysis
is generalized to the case of intersecting D-branes at general
angles \ct{bdl}. Since the rotational symmetry is reduced to $\so \times \sop$,
there are only two kinds of supersymmetric intersection at general
angles. One is generated by
$SU(2) \subset SO(4)$ or $\sop$ and the other is generated by $SU(2) \times
SU(2) \subset \so \times \sop$. It turns out that the former case further
breaks the supersymmetry by half after rotation while the latter
does by three quarter like in the flat spacetime. In Sec. 4,
remaining supersymmetries of intersecting D-branes are identified
by finding conserved worldsheet supercurrents consistent with open
string boundary conditions \ct{kim}. We summarize our results for the
unbroken supersymmetries in Tables 1-2. In section 5, the explicit
worldhseet expressions of conserved supercharges and their
supersymmetry algebras are presented. In section 6, we briefly
review our results and address some other issues. In Appendix A,
we present calculational details of the supersymmetry algebra for
a specific example, $D_-3 -D_-5$ intersection, to illustrate
nontrivial aspects of the derivation.

%%%%%%%%%%%%%%%%%%%%%%%%%%%%%%%%%%%%%%%%%%%%%%%%%%%%%%%%%%%%%%%%%%%%%%
\section{$p-q$ Strings in A Plane Wave Background}
%%%%%%%%%%%%%%%%%%%%%%%%%%%%%%%%%%%%%%%%%%%%%%%%%%%%%%%%%%%%%%%%%%%%%%

The Green-Schwarz light-cone action in the plane wave background \eq{pp-metric}
describes eight free massive bosons and fermions. In the
light-cone gauge, $X^+=\tau$, the action is given by
\be \la{gs-action}
S = \frac{1}{2\pi \a^\prime p^+} \int d\tau \int_{0}^{2\pi \a^\prime |p^+|}
d\sigma \Bigl[ \half \p_+X_I \p_-X_I - \half \mu^2 X_I^2
- i \bar{S}(\rho^A \p_A - \mu \Pi)S \Bigr]
\ee
where $\p_\pm = \p_\tau \pm \p_\sigma$.
In this paper we will
use the notation and the convention in \ct{kim}.
The equations of motion following from the action \eq{gs-action}
take the form
\bea \la{eom-boson}
&& \p_+\p_-X^I + \mu^2 X^I =0, \\
\la{eom-fermion}
&& \p_+ S^1 - \mu \Pi S^2 =0, \qquad \p_-S^2 + \mu \Pi S^1 =0.
\eea

We use the following form for $\gamma^I$
\be \la{8d-gamma}
\gamma^I = \pmatrix{ 0 & \gamma^I_{a\ad} \cr
                       \tilde{\gamma}^I_{\ad a} & 0 }
\ee
where $\tilde{\gamma}^I_{\ad a}=({\gamma^I}^T)_{\ad a}$
and take the $SO(8)$ chirality matrix as
\be \la{gamma9}
\gamma=\pmatrix{ 1_{8} & 0 \cr
                       0 & -1_{8} }.
\ee
In what follows, we assume that the spinors $S^A(\tau,\sigma), \; A=1,2,$
are positive chiral
fermions, $\gamma S^A = S^A$, of the form
\be \la{so8-spinor}
S_\a^A =\left(%
\begin{array}{c}
  S_a^A \\
  0 \\
\end{array}%
\right),
\ee
where $\a = 1, \cdots, 16$ and $a=1, \cdots, 8$.

Consider an open string stretched between $Dp$-brane and
$Dq$-brane in the plane wave background \eq{pp-metric}.
The open string action is just defined by the action \eq{gs-action}
with string length $\a = 2\a^\prime p^+$ imposed with
appropriate boundary conditions on each end of the open string.
For longitudinal coordinates $X^r$ on D-branes, we impose the
Neumann boundary condition
\be \la{nbc-boson}
\p_\sigma X^r|_{\p\Sigma}=0,
\ee
while for transverse coordinates $X^{r^\prime}$ we have the Dirichlet boundary
condition
\be \la{dbc-boson}
\p_\tau X^{r^\prime}|_{\p\Sigma}=0.
\ee
The fermionic coordinates also have to satisfy the following
boundary condition at each end of the open string
\ct{lambert}
\be \la{bc-fermion}
(S^1-\Omega_0 S^2)|_{\sigma=0}=0, \qquad
(S^1- \Omega_\pi S^2)|_{\sigma=\pi \a}=0,
\ee
where the matrix $\Omega_\theta =(\Omega_0, \Omega_\pi)$ is the products of
$\gamma$-matrices along worldvolume directions and,
depending on the type of D-branes, $D_\pm$-branes, satisfies
\be \la{dpm}
D_-: \Pi\Omega_\theta \Pi\Omega_\theta = -1, \qquad
D_+: \Pi\Omega_\theta\Pi\Omega_\theta =1.
\ee
The condition \eq{dpm} fairly
restricts the possible D-branes and their polarization
\ct{dabholkar,sken-tayl,bmz,skenderis1,gaberdiel}.
We use the notation $(+,-,m,n)$ \ct{sken-tayl} to indicate a brane
wrapping the light-cone coordinates $X^\pm$, $m$ coordinates that used to be $AdS_5$
coordinates before the Penrose limit, and $n$ coordinates that
used to be $S^5$ coordinates. Here we list the allowed choices:
\bea \la{d-brane}
&& D3: (+,-,m,n)\; =\;(+,-,2,0),\; (+,-,0,2), \xx
&& D5: (+,-,m,n)\; =\;(+,-,3,1),\; (+,-,1,3), \\
&& D7: (+,-,m,n)\; =\;(+,-,4,2),\; (+,-,2,4) \nonumber
\eea
for $D_-$-branes and
\bea \la{d+brane}
&& D1: (+,-,m,n)\; =\;(+,-,0,0), \xx
&& D3: (+,-,m,n)\; =\;(+,-,1,1), \xx
&& D5: (+,-,m,n)\; =\;(+,-,4,0),\; (+,-,2,2),\;(+,-,0,4), \\
&& D7: (+,-,m,n)\; =\;(+,-,3,3), \xx
&& D9: (+,-,m,n)\; =\;(+,-,4,4) \nonumber
\eea
for $D_+$-branes.

Now we will analyze the detailed statics of an open string
stretched between $Dp$-brane and $Dq$-brane in the
plane wave background \eq{pp-metric} from which we will determine the
residual supersymmetries of the D-brane configurations. In this
section we will first consider parallel D-branes and intersecting
D-branes at right angles. More general supersymmetric
intersections will be discussed in section 3.

%%%%%%%%%%%%%%%%%%%%%%%%%%%%%%%%%%%%%%%%%%%%%%%%%%%%%%%%%%%%%%%%%%%%%%
\subsection{$D_- - D_-$ Brane Configurations}
%%%%%%%%%%%%%%%%%%%%%%%%%%%%%%%%%%%%%%%%%%%%%%%%%%%%%%%%%%%%%%%%%%%%%%

We consider an open string intervened between parallel or
orthogonally intersecting D-branes.
The coordinates $X^I(\tau, \sigma)$ of a $p-q$ string
can be partitioned into four sets, NN, DD, ND, and DN, according
to whether the coordinate $X^I$ has Neumann (N) or Dirichlet (D)
boundary condition at each end. We first present the mode
expansion of the bosonic coordinates $X^I(\tau, \sigma)$ for the four
possible boundary conditions\footnote{In the following we will use indices
 $(r, s, \cdots), \;(r^\prime, s^\prime,\cdots),
\; (i,j,\cdots)$, and $(i^\prime,j^\prime, \cdots)$
for NN, DD, ND, and DN coordinates, respectively. We will also use
the subscripts $(n, m, \cdots) \in {\bf Z}$ for integer modes of string oscillators
while $(\k, \lambda, \cdots) \in {\bf R}$ for general real number modes.}:
\bea \la{mode-x}
&& \mbox{NN}:\; X^r(\tau, \sigma)=\cos\mu\tau x_0^r
+ \sin\mu\tau \frac{p_0^r}{\mu} +i\sum_{n
\neq 0}\frac{\a_n^{r}}{\omega_n} e^{-i\omega_n \tau}
\cos\frac{n\sigma}{|\a|}, \xx
&& \mbox{DD}:\; X^{r^\prime}(\tau, \sigma)= x_1^{r^\prime} \cosh \mu\sigma
 + \frac{x_2^{r^\prime}- x_1^{r^\prime} \cosh \pi \mu|\a|}{\sinh \pi \mu|\a|}
 \sinh \mu \sigma +
\sum_{n \neq 0}\frac{\a_n^{r^\prime}}{\omega_n} e^{-i\omega_n \tau}
\sin\frac{n\sigma}{|\a|}, \xx
&& \mbox{ND}:\; X^{i}(\tau, \sigma)= \frac{x_2^{i} \cosh \mu \sigma}{\cosh \pi \mu|\a|}
+ i \sum_{\k \in \bfz + \half}\frac{\a_\k^{i}}{\omega_\k} e^{-i\omega_\k \tau}
\cos \frac{\k\sigma}{|\a|}, \xx
&& \mbox{DN}:\; X^{i^\prime}(\tau, \sigma)= \frac{x_1^{i^\prime}
\cosh \mu (\sigma-\pi|\a|)}{\cosh \pi \mu|\a|}
+ \sum_{\k \in \bfz + \half}\frac{\a_\k^{i^\prime}}{\omega_\k} e^{-i\omega_\k \tau}
\sin\frac{\k\sigma}{|\a|},
\eea
where $x^I_1$ and $x_2^I$ denote the transverse positions of
$Dp$-brane and $Dq$-brane, respectively and
\be \la{omega}
\omega_\nu = \mbox{sgn}(\nu) \sqrt{\mu^2 + \nu^2/\a^2},
\quad \mbox{for} \; \nu = n, \k.
\ee
The commutation relations for the modes in Eq. \eq{mode-x}
are given by
\bea \la{comm-x}
&& [x_0^r, p_0^s]=i\delta^{rs}, \xx
&& [\a_n^{I},\a_m^{J}]= \omega_n
\delta_{m+n,0} \delta^{IJ}, \qquad
[\a_\k^{I},\a_\lambda^{J}]= \omega_\k
\delta_{\k + \lambda,0} \delta^{IJ}.
\eea

We are mainly interested in a supersymmetric intersection. In
particular, the dynamical supersymmetry transformation is given by
\be \la{dyn-susytr}
\delta_\epsilon X^I = \frac{1}{\sqrt{2p^+}} \bar{\epsilon}^A
\gamma^I S^A.
\ee
Hence, we take an appropriate combination of spinor
fields $\xi^A(\tau, \sigma)$ with integer modes and
$\eta^A(\tau,\sigma)$ with half-integer modes
to be compatible with the bosonic case \eq{mode-x}:
\bea \la{spinor-s}
&& S^1(\tau,\sigma)=\left\{%
\begin{array}{ll}
    I_+ \xi^1(\tau,\sigma) + I_- \eta^1(\tau,\sigma), & \hbox{for A-type;} \\
    I_- \xi^1(\tau,\sigma) + I_+ \eta^1(\tau,\sigma), & \hbox{for B-type,} \\
\end{array}%
\right. \xx
&& S^2(\tau,\sigma)=I_+ \xi^2(\tau,\sigma) + I_-
\eta^2(\tau,\sigma),
\eea
where $I_+$ and $I_-$ are $16 \times 16$ matrices satisfying
\be \la{inik}
I_+ + I_- =1, \quad  I_+ I_- = 0, \quad I_+^2 = I_+,
\quad I_-^2 =I_-.
\ee
The condition \eq{inik} simply states that one has to pick up only
eight components from the two $SO(8)$ chiral spinors $\xi^A(\tau,\sigma)$
and $\eta^A(\tau,\sigma)$ to give an $SO(8)$ chiral spinor $S^A(\tau,\sigma)$.
For the reason explained later, the A-type solution is for
$|p-q|=0,4,8$ in $Dp-Dq$ brane configurations while the B-type
solution for $|p-q|=2,6$. We take the spinors $\xi^A(\tau,\sigma)$
and $\eta^A(\tau,\sigma)$ as the solution of the equation of motion
\eq{eom-fermion} satisfying the boundary condition \eq{bc-fermion}
at $\sigma =0$ \ct{kim}:
\bea \la{open-fermion}
&& \xi^1 (\tau, \sigma)= \cos \mu\tau S_0
- \sin\mu\tau \Omega_0 \Pi S_0
+ \sum_{n \neq 0}c_n(\varphi_n^1(\tau, \sigma) \Omega_0 S_n
+i \rho_n \varphi_n^2(\tau, \sigma)\Pi S_n), \xx
&& \xi^2 (\tau, \sigma)= \cos \mu\tau \Omega^T_0 S_0
- \sin\mu\tau \Pi S_0
+ \sum_{n \neq 0}c_n(\varphi_n^2(\tau, \sigma) S_n
- i \rho_n \varphi_n^1(\tau, \sigma)\Pi \Omega_0 S_n), \xx
&& \eta^1 (\tau, \sigma)= \sum_{\k \in \bfz + \half}c_\k(\varphi_\k^1(\tau, \sigma)
\Omega_0 S_\k +i \rho_\k \varphi_\k^2(\tau, \sigma)\Pi S_\k), \xx
&& \eta^2 (\tau, \sigma)= \sum_{\k \in \bfz + \half}c_\k (\varphi_\k^2(\tau, \sigma)
S_\k - i \rho_\k \varphi_\k^1(\tau, \sigma)\Pi \Omega_0 S_\k),
\eea
where the basis functions $\varphi_\nu^{1,2}(\tau, \sigma)$ are
defined by
\be \la{basis}
\varphi_\nu^1(\tau, \sigma)=e^{-i(\omega_\nu \tau -
\frac{\nu}{|\a|}\sigma)}, \qquad \varphi_\nu^2(\tau, \sigma)=e^{-i(\omega_\nu \tau
+ \frac{\nu}{|\a|}\sigma)}
\ee
and
\be \la{omega-etc}
\omega_\nu= \mbox{sgn}(\nu) \sqrt{\mu^2 + \nu^2/\a^2}, \quad \rho_\nu=
\frac{\omega_\nu-\nu/|\a|}{\mu}, \quad c_\nu= \frac{1}{\sqrt{1+\rho_\nu^2}}.
\ee
Here $\nu$ is either integer $n$ or half-integer $\k$.
The commutation relations for the modes in \eq{open-fermion} are given by
\be \la{comm-s}
\{S_n^{a}, S_m^{b} \} = \frac{1}{4} \delta_{n+m,0}
\delta^{ab}, \qquad \{S_\k^{a}, S_\lambda^{b} \} = \frac{1}{4} \delta_{\k+\lambda,0}
\delta^{ab}.
\ee

At $\sigma = \pi |\a|$, the spinors satisfy the following
relations
\be \la{spinor-pi}
\xi^1(\tau, \sigma = \pi |\a|)= \Omega_0 \xi^2(\tau, \sigma = \pi
|\a|), \qquad \eta^1(\tau, \sigma = \pi |\a|)= - \Omega_0 \eta^2(\tau, \sigma = \pi
|\a|).
\ee
The boundary conditions \eq{bc-fermion} for the spinors $S^A(\tau,\sigma)$
in Eq. \eq{spinor-s} require the following property for the projection
matrices $I_\pm$:
\bea \la{bc-i+i-}
&& \Omega_0 I_\pm = \left\{%
\begin{array}{ll}
    I_\pm \Omega_0, & \hbox{for A-type;} \\
    I_\mp \Omega_0, & \hbox{for B-type,} \\
\end{array}%
\right. \quad \mbox{at}\; \sigma = 0, \xx
&& \Omega_0^T \Omega_\pi I_\pm = \pm I_\pm,
\quad \mbox{at}\; \sigma = \pi |\a|.
\eea
Eq. \eq{bc-i+i-} can be satisfied only if
the matrix $\Omega_0^T \Omega_\pi$ is symmetric, i.e.,
\be \la{symm-condition}
(\Omega_0^T \Omega_\pi)^T = \Omega_0^T \Omega_\pi.
\ee
The matrices $I_+$ and $I_-$ can be solved to give
\be \la{inik-sol}
I_+ =  \half(1 + \Omega_0^T \Omega_\pi), \qquad
I_- =  \half(1 - \Omega_0^T \Omega_\pi).
\ee

Note that $\Omega^T_\theta = - \Omega_\theta$ for $D3$- and $D7$-branes,
but $\Omega^T_\theta = \Omega_\theta$ for $D5$-branes and thus
\be \la{ab-type}
\Omega_0^T \Omega_\pi = \left\{%
\begin{array}{ll}
    \Omega_0 \Omega_\pi^T, & \hbox{for A-type;} \\
    - \Omega_0 \Omega_\pi^T, & \hbox{for B-type.} \\
\end{array}%
\right.
\ee
Using Eq. \eq{ab-type}, we get useful identities:
\bea \la{identity-a}
&& \Pi I_\pm = I_\pm \Pi, \quad \Omega_\theta I_\pm = I_\pm \Omega_\theta,
\quad I_\pm \Omega_0 = \pm  I_\pm
\Omega_\pi, \quad \mbox{for A-type;} \\
\la{identity-b}
&& \Pi I_\pm = I_\mp \Pi, \quad \Omega_\theta I_\pm = I_\mp \Omega_\theta,
\quad I_\pm \Omega_0 = \mp  I_\pm
\Omega_\pi, \quad \mbox{for B-type}.
\eea
On can easily see that the spinors in Eq. \eq{spinor-s} satisfy
the equations of motion \eq{eom-fermion} and the boundary conditions \eq{bc-fermion}.

Note that $\Omega_0^T \Omega_\pi$ consists of products of $\gamma$-matrices along the
ND and DN directions. Since $(\Omega_0^T \Omega_\pi)^2=1$ and $\Tr(\Omega_0^T \Omega_\pi)=0$
for $\Omega_0^T \Omega_\pi \neq \pm 1$, there can
be only three kinds of possibility:
\be \la{omega-sol}
\Omega_0^T \Omega_\pi= \left\{%
\begin{array}{ll}
    \pm 1, & \hbox{$\sharp_{ND}=0$}, \\
    \pm \gamma, & \hbox{$\sharp_{ND}=8$}, \\
    \pm \pmatrix{
  \Xi & 0 \cr
  0 & \pm \Xi \cr}, & \hbox{$\sharp_{ND}=4$}, \\
\end{array}%
\right.
\ee
where
\be \la{Xi}
\Xi = \pmatrix{
  1_4 & 0 \cr
  0 & -1_4 \cr},
\ee
and $\sharp_{ND}$ denotes the total number of ND and DN directions.

The case $\Omega_0^T \Omega_\pi=1$ corresponds to
parallel $Dp$-branes while the case $\Omega_0^T
\Omega_\pi=-1$ corresponds to $Dp$-anti-$Dp$ branes, but
the cases $\Omega_0^T \Omega_\pi=\pm \gamma$ and $\Omega_0^T
\Omega_\pi=\pm \Xi$ correspond to $Dp-Dq$ or $Dp$-anti-$Dq$
branes with $\sharp_{ND}=8$ and $\sharp_{ND}=4$, respectively.
Note that the B-type branes allow only the $\sharp_{ND}=4$ case as seen
from the list in \eq{d-brane}.
We will show in section 4 by deriving conserved worldsheet
supercurrents that the brane configurations in \eq{omega-sol} preserve
the same amount of supersymmetries as in flat spacetime as long as
two branes are at origin.

%%%%%%%%%%%%%%%%%%%%%%%%%%%%%%%%%%%%%%%%%%%%%%%%%%%%%%%%%%%%%%%%%%%%%%
\subsection{$D_+ - D_+$ Brane Configurations}
%%%%%%%%%%%%%%%%%%%%%%%%%%%%%%%%%%%%%%%%%%%%%%%%%%%%%%%%%%%%%%%%%%%%%%

We first analyze the mode expansions of bosons. $D_+ - D_+$ brane
configurations where both branes have no worldvolume fluxes have
the same mode expansion as the $D_- - D_-$ case. We now consider
the cases in which at least one of two $D_+$-branes is a
$D5$-brane with a worldvolume flux. In this case the $D5$-branes
of type $(+,-,4,0)$ or $(+,-,0,4)$ with a Born-Infeld flux satisfy
the modified Neumann boundary condition \ct{gaberdiel,kim}:
\be \la{+d5-nbc}
(\p_\sigma X^r - \mu X^r)|_{\p \Sigma}=0,
\qquad \forall \; r \in \;\mbox{N}.
\ee
First we consider the brane configuration consisting of a
$Dp$-brane
with no worldvolume flux at $\sigma =0$ and a $D5$-brane
with a worldvolume flux at $\sigma = \pi|\a|$.
In this brane configuration the bosonic
coordinates with NN, DD, ND, and DN boundary conditions have the
following mode expansions, respectively:
\bea \la{mode-d5d+}
&& \mbox{NN}:\; X^r(\tau, \sigma)=i\sum_{\k} \frac{\a_\k^{r}}{\omega_\k}
 e^{-i\omega_\k \tau}
\cos\frac{\k \sigma}{|\a|}, \xx
&& \mbox{DD}:\; X^{r^\prime}(\tau, \sigma)= x_1^{r^\prime} \cosh \mu\sigma
 + \frac{x_2^{r^\prime}- x_1^{r^\prime} \cosh \pi \mu|\a|}{\sinh \pi \mu|\a|}
 \sinh \mu \sigma +
\sum_{n \neq 0}\frac{\a_n^{r^\prime}}{\omega_n} e^{-i\omega_n \tau}
\sin\frac{n\sigma}{|\a|}, \xx
&& \mbox{ND}:\; X^{i}(\tau, \sigma)= \frac{x_2^{i} \cosh \mu \sigma}{\cosh \pi \mu|\a|}
+ i \sum_{\k \in {\bf Z} + \half} \frac{\a_\k^{i}}{\omega_\k} e^{-i\omega_\k \tau}
\cos \frac{\k\sigma}{|\a|}, \xx
&& \mbox{DN}:\; X^{i^\prime}(\tau, \sigma)= x_1^{i^\prime} e^{\mu \sigma}
+ \sum_{\lambda}\frac{\a_\lambda^{i^\prime}}{\omega_\lambda} e^{-i\omega_\lambda \tau}
\sin\frac{\lambda \sigma}{|\a|},
\eea
where $\k$ and $\lambda$ in NN and DN coordinates are generically irrational numbers
determined by the equations
\be \la{d5-trans-eq}
e^{2\pi i \k} = \frac{\k - i \mu |\a|}{\k + i \mu |\a|},
\qquad e^{2\pi i \lambda} = - \frac{\lambda - i \mu |\a|}
{\lambda + i \mu |\a|},
\ee
respectively. The frequencies $\omega_{\kappa,\lambda}$ in the NN and DN coordinates
are defined as in Eq. \eq{omega-etc} with $\kappa$ and $\lambda$ satisfying the
equations \eq{d5-trans-eq}.
It is obvious that there are infinitely many
solutions in Eq. \eq{d5-trans-eq} for $\k$ and $\lambda$.
Also, if $\k \,(\lambda)$ is a solution to the first (second)
equation, then $-\k \,(-\lambda)$ is also a solution. The reality
condition for NN and DN coordinates is thus satisfied only if
$(\a_\k^{i})^\dagger = \a_{-\k}^{i}$ and
$(\a_{\lambda}^{i^\prime})^\dagger = \a_{-\lambda}^{i^\prime}$.
Note that even NN coordinates do not have zero modes
since $\k =0$ is not a solution of the first equation in Eq.
\eq{d5-trans-eq}. This is due to different behaviors of zero modes
on $D_+$-branes because the behavior on $D_+$-brane without flux
has an oscillatory behavior around some fixed position while that
on $D_+$ brane with flux tends to move
with constant velocity \ct{kim}.
DN coordinates do not contain any zero modes either
since a zero mode solution with $\lambda=0$
in the DN directions is identically zero
although it is a trivial solution of Eq. \eq{d5-trans-eq}.
The commutation relations for the modes in Eq. \eq{mode-d5d+}
are identical to the case \eq{mode-x}.

Next we consider the configurations of $(+,-,4,0)$ and $(+,-,0,4)$
branes with worldvolume fluxes altogether \ct{kim}. In order to find the mode expansion
in this case, it is convenient to introduce new coordinates
$\widetilde{X}^I \equiv e^{-\mu\sigma}X^I$. The new coordinates
$\widetilde{X}^I(\tau, \sigma)$ then satisfy the modified equation of
motion
\be \la{eom-new}
(\p_\tau^2-\p_\sigma^2-2 \mu \p_\sigma)\widetilde{X}^I = 0
\ee
and the usual boundary conditions
\be \la{bc-new}
\p_\sigma \widetilde{X}^r|_{\p \Sigma} = 0, \qquad
\p_\tau \widetilde{X}^{r^\prime}|_{\p \Sigma} = 0.
\ee
The mode expansions of the bosonic coordinates
$\widetilde{X}^I(\tau, \sigma)$ and thus $X^I(\tau, \sigma)$
are solved conveniently by a method called separation of variables
for NN, DD, ND, and DN boundary conditions:
\bea \la{mode-d+}
&& \mbox{NN}:\; X^r(\tau, \sigma)=\sqrt{\frac{2 \pi \mu |\a|}{e^{2 \pi \mu
|\a|}-1}} (x_0^r + p_0^r \tau) e^{\mu \sigma}+ \frac{i}{2} \sum_{n \neq 0}
\frac{\a_n^r}{\omega_n} \Bigl( \varphi_n^1(\tau,\sigma) +
\frac{n + i \mu |\a|}{n - i \mu |\a|} \varphi_n^2(\tau,\sigma) \Bigr), \xx
&& \mbox{DD}:\; X^{r^\prime}(\tau, \sigma)= x_1^{r^\prime} \cosh \mu\sigma
 + \frac{x_2^{r^\prime}- x_1^{r^\prime} \cosh \pi \mu|\a|}{\sinh \pi \mu|\a|}
 \sinh \mu \sigma +
\sum_{n \neq 0}\frac{\a_n^{r^\prime}}{\omega_n} e^{-i\omega_n \tau}
\sin\frac{n\sigma}{|\a|}, \xx
&& \mbox{ND}:\; X^{i}(\tau, \sigma)= x_2^{i}
e^{\mu(\sigma - \pi|\a|)}
+ \frac{i}{2} \sum_{\k} \frac{\a_\k^{i}}{\omega_\k}
\Bigl( \varphi_\k^1(\tau,\sigma)- e^{2\pi i \k}
\varphi_\k^2(\tau,\sigma) \Bigr), \xx
&& \mbox{DN}:\; X^{i^\prime}(\tau, \sigma)= x_1^{i^\prime} e^{\mu\sigma}
+ \frac{1}{2i} \sum_{\lambda} \frac{\a_{\lambda}^{i^\prime}}{\omega_\lambda}
\Bigl( \varphi_{\lambda}^1(\tau,\sigma)-
\varphi_{\lambda}^2(\tau,\sigma) \Bigr),
\eea
where the mode numbers $\k$ and $\lambda$ are determined by the equations
\be \la{trans-eq}
e^{2\pi i \k} = - \frac{\k + i \mu |\a|}{\k - i \mu |\a|},
\qquad e^{2\pi i \lambda} = - \frac{\lambda - i \mu |\a|}{\lambda + i \mu
|\a|},
\ee
respectively. The basis functions $\varphi^{1,2}_{\kappa,\lambda} (\tau,\sigma)$
are defined as in Eq. \eq{basis} with $\kappa$ and $\lambda$ satisfying the
equations \eq{trans-eq}.

The mode expansion of the spinor field can be determined by
following exactly the same procedure as in the previous
subsection. We take a combination of spinor
fields $\xi^A(\tau, \sigma)$ with integer modes and spinor fields
$\eta^A(\tau,\sigma)$ with ${\bf R}$-modes:
\bea \la{spinor-d+}
&& S^1(\tau,\sigma)=\left\{%
\begin{array}{ll}
    I_+ \xi^1(\tau,\sigma) + I_- \eta^1(\tau,\sigma), & \hbox{for A-type;} \\
    I_- \xi^1(\tau,\sigma) + I_+ \eta^1(\tau,\sigma), & \hbox{for B-type,} \\
\end{array}%
\right. \xx
&& S^2(\tau,\sigma)=I_+ \xi^2(\tau,\sigma) + I_-
\eta^2(\tau,\sigma),
\eea
where
\bea \la{fermion-d+d+}
&& \xi^1 (\tau, \sigma)=\cosh \mu\sigma S_0
+ \sinh \mu\sigma \Omega_0 \Pi S_0
+ \sum_{n \neq 0}c_n(\varphi_n^1(\tau,\sigma) \Omega_0 \widetilde{S}_n
+i \rho_n \varphi_n^2(\tau,\sigma)\Pi S_n), \xx
&& \xi^2 (\tau, \sigma)= \cosh \mu\sigma \Omega_0^T S_0
+ \sinh \mu\sigma \Pi S_0
+ \sum_{n \neq 0}c_n(\varphi_n^2(\tau,\sigma) S_n
- i \rho_n \varphi_n^1(\tau,\sigma)\Pi \Omega_0 \widetilde{S}_n), \xx
&& \eta^1 (\tau, \sigma)= \sum_{\k}c_\k( \varphi_\k^1(\tau,\sigma)
\Omega_0 \widetilde {S}_\k
+ i \rho_\k \varphi_\k^2(\tau,\sigma)\Pi S_\k ), \xx
&& \eta^2 (\tau, \sigma)= \sum_{\k}c_\k (\varphi_\k^2(\tau,\sigma) S_\k
- i \rho_\k \varphi_\k^1(\tau,\sigma)\Pi \Omega_0 \widetilde{S}_\k ).
\eea
Similarly, we call A-type branes for $|p-q|=0,4,8$ in $Dp-Dq$
brane configurations while B-type branes for $|p-q|=2,6$.
The projection matrices $I_\pm$ are equally given by Eq.
\eq{inik-sol} and thus satisfy the identities, Eqs. \eq{identity-a}
and \eq{identity-b}. Therefore only three kinds of possibility in
\eq{omega-sol} are allowed.

In Eq. \eq{fermion-d+d+}, we introduced
$c_\kappa$ and $\rho_\kappa$ defined as in Eq. \eq{omega-etc}
with $\nu$ replaced by $\kappa$ being a solution of either the first
or the second equation in Eq. \eq{trans-eq} and
\be \la{sn}
\widetilde{S}_n = \frac{1}{\omega_n}\Bigl(\frac{n}{|\a|}
- i \mu \Pi  \Omega_0 \Bigr) S_n, \qquad
\widetilde{S}_\k = \frac{1}{\omega_\k}\Bigl(\frac{\k}{|\a|}
- i \mu \Pi  \Omega_0 \Bigr) S_\k.
\ee
One can see that a zero mode solution in $\eta^A(\tau,\sigma)$
with $\k =0$ is identically cancelled as expected.

In order to see that Eq. \eq{spinor-d+} satisfies the boundary condition \eq{bc-fermion},
it is more convenient to decompose the spinors $S^A_\k$ into
eigenspinors of $\Pi\Omega_0$ by defining
\be \la{dec-d+}
S_\k^\pm = \half(1 \pm \Pi\Omega_0) S_\k.
\ee
The spinors then have the following property
\be \la{prop-d+}
\Pi\Omega_0 S_\k^\pm = \pm S_\k^\pm, \qquad \Pi S_\k^\pm = \pm
\Omega_0 S_\k^\pm.
\ee
Using this property, $\eta^A(\tau, \sigma)$ in Eq.
\eq{fermion-d+d+} can be rewritten as
\bea \la{rew-eta1}
\eta^1 (\tau, \sigma) &=& \sum_{\k}c_\k \Bigl( \frac{\frac{\k}{|\a|} - i \mu}{\omega_\k}
\varphi_\k^1(\tau,\sigma) + i \rho_\k \varphi_\k^2(\tau,\sigma)\Bigr)
\Omega_0 S_\k^+ \xx
&& + \sum_{\k}c_\k \Bigl( \frac{\frac{\k}{|\a|} + i \mu}{\omega_\k}
\varphi_\k^1(\tau,\sigma) - i \rho_\k \varphi_\k^2(\tau,\sigma)\Bigr)
\Omega_0 S_\k^-, \\
\la{rew-eta2}
\eta^2 (\tau, \sigma) &=& \sum_{\k}c_\k \Bigl( \varphi_\k^2(\tau,\sigma)
-i \frac{\rho_k}{\omega_k}(\frac{\k}{|\a|} - i \mu) \varphi_\k^1(\tau,\sigma) \Bigr)
S_\k^+  \xx
&& + \sum_{\k}c_\k \Bigl( \varphi_\k^2(\tau,\sigma)
+ i \frac{\rho_k}{\omega_k}(\frac{\k}{|\a|} + i \mu) \varphi_\k^1(\tau,\sigma)
\Bigr) S_\k^-.
\eea
Using the identities,
\be \la{2id}
\frac{\frac{\k}{|\a|} \mp i \mu}{\omega_\k} \pm i \rho_\k
= 1 \mp i \frac{\rho_\k}{\omega_\k}(\frac{\k}{|\a|} \mp i \mu),
\ee
one can easily see that Eq. \eq{spinor-d+}
satisfies the boundary conditions \eq{bc-fermion} if
the mode number $\k$ of $S_\k^+$ satisfies
the first equation in \eq{trans-eq} while the mode number $\k$ of $S_\k^-$
does the second equation. The commutation relations between the
modes read as
\bea \la{com-rel-d+d+}
&& \{S_0^{a}, S_0^{b} \} = \frac{\pi \mu |\a|}{4\sinh \pi \mu |\a|}
\Bigl(\delta^{ab}\cosh\pi\mu|\a|-(\Omega_0 \Pi)^{ab}\sinh\pi\mu|\a|
\Bigr), \xx
\la{d+1n0}
&& \{S_n^{a}, S_m^{b} \} = \frac{1}{4} \delta_{n+m,0}
\delta^{ab}, \qquad (n,m \neq 0), \\
&& \{ S_\k ^{\pm a}, S_\lambda^{\pm b} \} = \frac{1}{4}
\delta_{\k + \lambda,0} P^{ab}_\pm, \quad
\mbox{where}\; P_\pm = \half (1 \pm \Pi\Omega_0). \nonumber
\eea

%%%%%%%%%%%%%%%%%%%%%%%%%%%%%%%%%%%%%%%%%%%%%%%%%%%%%%%%%%%%%%%%%%%%%%
\subsection{$D_\mp - D_\pm$ Brane Configurations}
%%%%%%%%%%%%%%%%%%%%%%%%%%%%%%%%%%%%%%%%%%%%%%%%%%%%%%%%%%%%%%%%%%%%%%

The mode expansion of bosonic coordinates $X^I(\tau, \sigma)$ in $D_\mp - D_\pm$
brane configurations without a worldvolume flux is exactly the
same as the previous $D_- - D_-$ case, which is
given by Eq. \eq{mode-x}. And, for $D_\mp - D_\pm$ brane configurations
where $D_+$-brane is a $D5$-brane with the worldvolume flux, the
mode expansion of bosonic coordinates is exactly the same as Eq. \eq{mode-d5d+}.

From the previous analysis, we have seen that it is crucial
that $\Omega_0^T \Omega_\pi$ is a symmetric matrix for the spinors
$S^A(\tau,\sigma)$ to satisfy the boundary conditions
\eq{bc-fermion}. If $\Omega_0^T \Omega_\pi$ were an
antisymmetric matrix and thus $(\Omega_0^T \Omega_\pi)^2 = -1$,
it would have all eigenvalues $\pm i$ and so the boundary
conditions \eq{bc-fermion} could not simultaneously be satisfied.
Furthermore the matrices $I_\pm$ would have complex
eigenvalues, so they could no longer be projection matrices. This
leads to an intriguing consequence: $(+,-,4,0)$- and
$(+,-,0,4)$-branes have no supersymmetric intersections with
$D_-$-branes since $\Omega_0^T \Omega_\pi$ in this case is always an
antisymmetric matrix and thus they cannot satisfy
the boundary conditions \eq{bc-fermion}
and $D_+1$-brane cannot have a supersymmetric intersection with
$D_-3$- and $D_-7$-branes.
We will consider only the cases satisfying $(\Omega_0^T \Omega_\pi)^T
= \Omega_0^T \Omega_\pi$.\footnote{
This is the same reason that we excluded nonsupersymmetric brane
configurations with $\sharp_{ND} = 2,6$.
Note that the boundary state formalism also meets a similar
situation since the guiding principle in the construction of
boundary states is the preservation of various
supersymmetries \ct{gg}.} Since the symmetric condition for the matrix
$\Omega_0^T \Omega_\pi$ excludes $D5$-branes with flux, the bosonic mode
expansion for the $D_\mp - D_\pm$ case becomes exactly equal to the $D_-
- D_-$ case.

There are the following identities for A,B-type D-branes:
\bea \la{d-d+a}
&& \Pi I_\pm = I_\mp \Pi, \quad \Omega_\theta I_\pm = I_\pm \Omega_\theta,
\quad I_\pm \Omega_0 = \pm  I_\pm
\Omega_\pi, \quad \mbox{for A-type;} \\
\la{d-d+b}
&& \Pi I_\pm = I_\pm \Pi, \quad \Omega_\theta I_\pm = I_\mp \Omega_\theta,
\quad I_\pm \Omega_0 = \mp  I_\pm
\Omega_\pi, \quad \mbox{for B-type}.
\eea
Note that a new feature arises in this case. Unlike as the $D_\pm - D_\pm$ cases,
the matrices $I_\pm$ change in different way passing the matrices
$\Pi$ and $\Omega_\theta$ as seen in Eqs. \eq{d-d+a} and \eq{d-d+b}.
Thus, any solution of the equations of
motion \eq{eom-fermion} for spinors cannot be simultaneously compatible
with the boundary condition \eq{bc-fermion}.
This immediately implies that the $D_\mp - D_\pm$ brane
intersections preserve no supersymmetry.\footnote{We thank a referee of Physical
Review ${\bf D}$ for drawing our attention to clarify this
problem.}

Let us briefly explain why it should be. First take a spinor of
the following form as usual:
\be \la{spinor-s1}
S^1(\tau,\sigma)=\left\{%
\begin{array}{ll}
    I_+ \xi^1(\tau,\sigma) + I_- \eta^1(\tau,\sigma), & \hbox{for A-type;} \\
    I_- \xi^1(\tau,\sigma) + I_+ \eta^1(\tau,\sigma), & \hbox{for B-type,} \\
\end{array}%
\right. \xx
\ee
where the spinors $\xi^1(\tau,\sigma)$ and $\eta^1(\tau,\sigma)$
are supposed to satisfy the equations of
motion \eq{eom-fermion} and the boundary condition \eq{bc-fermion} at $\sigma=0$.
In order for the spinor $S^1(\tau,\sigma)$ to satisfy the equation
of motion, $\p_+ S^1 - \mu \Pi S^2=0$, the spinor
$S^2(\tau,\sigma)$ is necessarily of the form
\be \la{spinor-s2}
S^2(\tau,\sigma)=I_- \xi^2(\tau,\sigma) + I_+
\eta^2(\tau,\sigma).
\ee
However, the spinors $S^A(\tau,\sigma)$ in Eqs.
\eq{spinor-s1}-\eq{spinor-s2} do not satisfy the boundary
condition at $\sigma=0$:
\bea \la{bc-s1s2}
\Omega_0 S^2(\tau,\sigma=0) &=& \left\{%
\begin{array}{ll}
    I_- \xi^1(\tau,\sigma=0) + I_+ \eta^1(\tau,\sigma=0), & \hbox{for A-type;} \\
    I_+ \xi^1(\tau,\sigma=0) + I_- \eta^1(\tau,\sigma=0), & \hbox{for B-type,} \\
\end{array}%
\right. \xx
&\neq& S^1(\tau,\sigma=0).
\eea

It is easily shown that the solution in Eqs. \eq{spinor-s1}-\eq{spinor-s2}
cannot be compatible with supersymmetries. As we mentioned at the
beginning of this subsection, NN and DD
coordinates in the $D_\mp - D_\pm$ brane intersection have integer modes
while ND and DN coordinates have half-integer modes. Thus, if the
dynamical supersymmetry were strictly preserved, the supersymmetry
transformation, for example, for an NN coordinate $X^r$ and an ND
coordinate $X^i$ would be of the form
\be \la{dyn-susy-tr}
\delta_\epsilon X^r = \frac{1}{\sqrt{2p^+}} \bar{\epsilon}^A
\gamma^r \xi^A, \quad \delta_\epsilon X^i = \frac{1}{\sqrt{2p^+}}
\bar{\epsilon}^A \gamma^i \eta^A,
\ee
because the spinors $\xi^A$ and $\eta^A$ are supposed to be
described by integer modes and half-integer modes, respectively.
In order to satisfy the supersymmetric transformation
\eq{dyn-susy-tr}, the spinor $S^2(\tau, \sigma)$ should be of the
form
\be \la{susy-spinor-s2}
S^2(\tau,\sigma)=I_+ \xi^2(\tau,\sigma) + I_-
\eta^2(\tau,\sigma).
\ee
To see this, first note that the constant spinors
$\epsilon^A$ in the supersymmetry transformation \eq{dyn-susy-tr}
satisfy
\be \la{e-spinor}
\epsilon^1 = \Omega_0 \epsilon^2, \quad \epsilon^1 = \Omega_\pi \epsilon^2
\ee
and thus do
\bea \la{e-eigen-s}
&& \epsilon^1=\left\{%
\begin{array}{ll}
    \Omega_0^T \Omega_\pi \epsilon^1,  & \hbox{for A-type;} \\
    - \Omega_0^T \Omega_\pi \epsilon^1,  & \hbox{for B-type,} \\
\end{array}%
\right. \xx
&& \epsilon^2=\Omega_0^T \Omega_\pi \epsilon^2.
\eea
Then it is easy to show, using Eqs. \eq{e-eigen-s}, \eq{gamma-omega-comm},
and \eq{gamma-inik}, that the supersymmetry transformation
\eq{dyn-susytr} reduces to Eq. \eq{dyn-susy-tr}.
Indeed this explains why the mode expansion for supersymmetric
intersecting D-branes necessarily takes the
form such as Eqs. \eq{spinor-s} and \eq{spinor-d+}.
The solution \eq{spinor-s2} therefore is contradictory to the
dynamical supersymmetry. The similar thing happens for the
kinematical supersymmetry since the preserved kinematical
supersymmetry has to satisfy both the equation of motion and the
boundary condition. Thus all $D_\mp - D_\pm$
brane intersections preserve no supersymmetry.

%%%%%%%%%%%%%%%%%%%%%%%%%%%%%%%%%%%%%%%%%%%%%%%%%%%%%%%%%%%%%%%%%%%%%%
\section{D-branes at Angles}
%%%%%%%%%%%%%%%%%%%%%%%%%%%%%%%%%%%%%%%%%%%%%%%%%%%%%%%%%%%%%%%%%%%%%%

In the previous section we considered only either parallel
D-branes or intersecting D-branes at right angles. In this section
we will consider D-branes intersecting at general angles \ct{bdl}.

In flat spacetime, it was shown in \ct{bdl} (see also \ct{ohta-town})
that the condition for D-branes intersecting at angles to preserve
supersymmetry is that the two branes should be related by an
$SU(N)$ subgroup in a space rotation group $SO(d)$. The type IIB
plane wave background
\eq{pp-metric} has the space rotation group $\so \times \sop$.
Under any rotation $R$,
the brane characterized by $\Omega$ is mapped to a brane described
by $\widehat{\Omega} = R^T \Omega R$. Indeed, one can consider two
kinds of rotation in the plane wave background as recently
emphasized in \ct{ggsns}. One is that $R$ is an element in $\so
\times \sop$. The other is that $R$ is an element in $SO(8)$ that
is not in $\so \times \sop$. We call the latter an oblique
rotation. Since $\so \times \sop$ rotation commutes with $\Pi$,
and hence,
\be \la{ro-d-d+}
\Pi \widehat{\Omega}\Pi\widehat{\Omega} = \left\{%
\begin{array}{ll}
    -1, & \hbox{for}\; \Omega \in D_-; \\
    +1, & \hbox{for} \; \Omega \in D_+, \\
\end{array}%
\right.
\ee
the image of any supersymmetric brane under a rotation $R \in \so \times
\sop$ therefore describes another supersymmetric D-brane of the
same kind. On the other hand it can be shown \ct{ggsns} that
an oblique rotation $R$ has to satisfy
\be \la{r-oblique}
R^4 = 1.
\ee
We will not discuss the oblique D-branes in detail in this paper.

We are interested in D-branes intersecting at general angle, but
preserving some fraction of supersymmetry. We will
consider only intersecting D-branes generated by a subgroup of $\so \times \sop$
rotation. As shown in \ct{bdl}, the condition that two branes intersecting at angles
preserve a common supersymmetry is that they should be related by
an $SU(N)$ rotation in the $\so \times \sop$ group. Since $\so \times \sop =
(SU(2)_L \times SU(2)_R)^2$, the $SU(N)$ factors in $\so \times
\sop$ group are only $SU(2)_L$ and $SU(2)_R$, which are self-dual and
anti-self-dual rotations in $SO(4)$'s, respectively. Since
rotations that preserve the worldvolume of a brane do not change
the resulting orientation of the brane, we restrict our attention
to those rotating a plane which lies in the Neumann and Dirichlet
directions of a brane.

One can see from the list in Eqs. \eq{d-brane} and
\eq{d+brane} that the supersymmetric rotation is possible only for
$D3$- and $D7$-branes among the $D_-$-branes and $(+,-,2,2)$-brane
in $D_+$-branes since for other branes a nontrivial rotation is
$U(1)$ rotation, so completely breaks the supersymmetry. We first
consider a $D3$-brane at first oriented along the $X^{1,3}$-axes,
for example, and then
rotated by the angle $\phi_1$ in the $X^1X^2$ plane and $\phi_2$ in
the $X^3X^4$ plane. (A rotated $D7$-brane can also be treated
similarly.) We define the complex coordinates $Z^1 = X^1 + i X^2$
and $Z^2 = X^3 + i X^4$. There can be two kinds of boundary
conditions on stretched open strings. The first one is of the
NN-DD type given by
\bea \la{nnbbc-ro-brane}
&& \mbox{Re}\, \p_\sigma Z^i|_{\sigma=0}=0=
\mbox{Im}\,Z^i|_{\sigma=0}, \xx
&& \mbox{Re}\, e^{i\phi_i} \p_\sigma Z^i|_{\sigma=\pi|\a|}= 0 =
\mbox{Im}\, e^{i\phi_i} Z^i|_{\sigma=\pi |\a|},
\eea
where $i=1,2$ and the mode expansion of complex bosonic
coordinates is
\be \la{mode-complex}
Z^i(\tau, \sigma)=i \sum_{n_i \in \bfz} \Bigl(
\frac{\a_{\k_i}^i}{\omega_{\k_i}} \varphi_{\k_i}^1(\tau, \sigma)
+ \frac{\widetilde{\a}_{\lambda_i}^i}{\omega_{\lambda_i}}
\varphi_{\lambda_i}^2(\tau, \sigma) \Bigr),
\ee
where $\k_i = n_i - \delta_i$ and $\lambda_i = n_i + \delta_i$ with
$0 \leq \delta_i = \phi_i/\pi \leq \half$.
The second one is of the DN-ND type given by
\bea \la{dnbbc-ro-brane}
&& \mbox{Im}\, \p_\sigma Z^i|_{\sigma=0}=0=
\mbox{Re}\, Z^i|_{\sigma=0}, \xx
&& \mbox{Re}\, e^{i\phi_i} \p_\sigma Z^i|_{\sigma=\pi|\a|}= 0 =
\mbox{Im}\, e^{i\phi_i} Z^i|_{\sigma=\pi |\a|}
\eea
and the mode expansion is
\be \la{dn-mode-complex}
Z^i(\tau, \sigma)= \sum_{n_i \in \bfz} \Bigl(
\frac{\a_{\k_i}^i}{\omega_{\k_i}} \varphi_{\k_i}^1(\tau, \sigma)
+ \frac{\widetilde{\a}_{\lambda_i}^i}{\omega_{\lambda_i}}
\varphi_{\lambda_i}^2(\tau, \sigma) \Bigr),
\ee
where $\k_i = n_i + \half - \delta_i$
and $\lambda_i = n_i - \half + \delta_i$. The basis functions
$\varphi^{1,2}_{\kappa_i,\lambda_i} (\tau,\sigma)$ and
the frequencies $\omega_{\kappa_i,\lambda_i}$ are defined as in Eqs. \eq{basis} and
\eq{omega-etc} with $\kappa_i$ and $\lambda_i$. The oscillators
have the reality conditions $\a_{-\k_i}^i = \widetilde{\a}_{\k_i}^{i \,\dagger}$
and $\widetilde{\a}_{-\lambda_i}^i = \a_{\lambda_i}^{i \,\dagger}$
and the commutation relation is
\be \la{comm-ro}
[\a_{\k_i}^{i}, \a_{\k_j}^{j \, \dagger}] = \half \omega_{\k_i}
\delta_{n_i,m_j}\delta^{ij}, \qquad
[\widetilde{\a}_{\lambda_i}^{i}, \widetilde{\a}_{\lambda_j}^{j \, \dagger}]
= \half \omega_{\lambda_i}
\delta_{n_i,m_j}\delta^{ij}.
\ee

First we consider intersecting $D_- - D_-^\prime$ branes at
angles, especially $D_-p - D3^\prime$ as a representative example,
where a rotated brane is indicated with the prime.
The boundary conditions for spinor fields $S^A(\tau,\sigma)$ are
now given by
\be \la{bc-ro-fermion}
(S^1-\Omega_0 S^2)|_{\sigma=0}=0, \qquad
(S^1- \widehat{\Omega}_\pi S^2)|_{\sigma=\pi \a}=0,
\ee
where
\be \la{omega-hat}
\widehat{\Omega}_\pi = \Omega_\pi R^2, \qquad \Omega_\pi = \gamma^{13}
\ee
and
\be \la{R}
R = e^{\half(\gamma^{12} \phi_1 + \gamma^{34} \phi_2)}.
\ee
When $[R, \Omega_0]=0$, e.g. $(+,-,4,2)$-brane, the boundary
condition \eq{bc-ro-fermion} can be rewritten as the ordinary
boundary condition \eq{bc-fermion} with respect to the rotated
spinors $S^{\prime A} = R S^{A}$. Since
the action \eq{gs-action} is invariant under $\so \times \sop$
rotations, the resulting supersymmetry is never changed. Thus we
will not consider such cases either.

One may take explicit
eigenvalues of the rotation $R$ in \eq{R} as follows:
\be \la{rep-R}
R = e^{i\sum_{i=1}^2 s_i \phi_i}
\ee
where $s_i = \pm \half$ are eigenvalues of $\frac{i}{2}\gamma^{12}$
and $\frac{i}{2}\gamma^{34}$
acting on spinor fields. Since the second boundary condition can be
rewritten as the form $(S^1- \Omega_0 (\Omega_0^T\Omega_\pi)
R^2 S^2)|_{\sigma=\pi \a}=0$, one can see that the boundary condition
becomes identical to the original boundary condition before the
rotation if $R^2 = 1$, namely, $\phi_1 = \phi_2$, self-dual rotation,
and $\phi_1 = - \phi_2$, anti-self-dual rotation. However
the latter condition essentially reduces the number of spinors
compatible with the boundary condition by half compared to the
original brane configuration before rotation. Thus we expect that,
when $R^2 =1$, the supersymmetry is also further reduced by half,
otherwise, the supersymmetry is completely broken. We will prove
this claim in section 4.

The spinors $S^A(\tau,\sigma)$ have the same form as Eq. \eq{spinor-s}
where the projection matrices $I_\pm$ are still given by Eq. \eq{inik-sol},
but the mode numbers are quite different from Eq. \eq{open-fermion}
due to the second boundary condition in Eq. \eq{bc-ro-fermion}:
\bea \la{rot-fermion}
&& \xi^1 (\tau, \sigma)= \xi^1_0 (\tau, \sigma) +
\sum_{\k} c_\k(\varphi_\k^1(\tau, \sigma) \Omega_0
S_\k +i \rho_\k \varphi_\k^2(\tau, \sigma)\Pi S_\k), \xx
&& \xi^2 (\tau, \sigma)= \xi^2_0 (\tau, \sigma) +
\sum_{\k} c_\k(\varphi_\k^2(\tau, \sigma)
S_\k - i \rho_\k \varphi_\k^1(\tau, \sigma)\Pi \Omega_0 S_\k), \xx
&& \eta^1 (\tau, \sigma)= \sum_{\lambda}c_\lambda(\varphi_\lambda^1(\tau, \sigma)
\Omega_0 S_\lambda +i \rho_\lambda \varphi_\lambda^2(\tau, \sigma)\Pi S_\lambda), \xx
&& \eta^2 (\tau, \sigma)= \sum_{\lambda}c_\lambda (\varphi_\lambda^2(\tau, \sigma)
S_\lambda - i \rho_\lambda \varphi_\lambda^1(\tau, \sigma)\Pi \Omega_0
S_\lambda),
\eea
where $\xi^A_0 (\tau, \sigma)$ are possible zero modes to be
determined later.
The first boundary condition in Eq. \eq{bc-ro-fermion} is
automatically satisfied due to the properties in Eqs.
\eq{identity-a} and \eq{identity-b}.
Let us now briefly explain what condition arises and
how to determine the mode numbers $\k$ and
$\lambda$ from the second boundary condition in Eq.
\eq{bc-ro-fermion}. A nontrivial requirement is that
$[R, I_\pm]=0$ or equivalently, $R \Omega_0 = \Omega_0 R^T$,
which means that
\be \la{ro-gamma-omega}
\{\gamma^{12}, \Omega_0 \} = \{ \gamma^{34}, \Omega_0 \} = 0,
\ee
since we already excluded the case $[R, \Omega_0]=0$ for the reason
explained above.
This requires that the brane charaterized by $\Omega_0$
has to span $X^1-X^3$ plane or $X^2-X^4$ plane. Then the second
boundary condition in Eq. \eq{bc-ro-fermion} reduces to the
following equations which determine the mode numbers $\k$ and
$\lambda$:
\bea \la{ro-mode-number}
&& e^{2\pi i \k} S_\k - i \rho_\k \Pi \Omega_0 S_\k =
R^2 (S_\k - i e^{2\pi i \k} \rho_\k \Pi \Omega_0 S_\k ), \xx
&& e^{2\pi i \lambda} S_\lambda
- i \rho_\lambda \Pi \Omega_0 S_\lambda = - R^2
(S_\lambda - i e^{2\pi i \lambda}
\rho_\lambda \Pi \Omega_0 S_\lambda ).
\eea
The equations
\eq{ro-mode-number} can be easily solved and the result is given by
\be \la{klam}
\k = n + \nu_a, \qquad \lambda = n - \half + \nu_a,
\qquad n \in \bfz,
\ee
where $\nu_a = \sum_{i=1}^2 s_i \delta_i$. Of course, the phases
$\nu_a \;(a=1,\cdots,8)$ depend on the eigenvalues $s_i$ of
the spinors $S_\k$ and $S_\lambda$ for given angles $\phi_i$, but the
details are not important in our context.

For the case $\phi_1 = \phi_2$ and $\phi_1 = - \phi_2$,
zero modes with $\k =0$ exist for $(2s_1,2s_2) = (\pm 1,\mp 1)$
and $(\pm 1,\pm 1)$, respectively and they are
of the same form as the zero modes in \eq{open-fermion}, but with
$S_0$ satisfying $R^2 S_0 = S_0$, i.e., $\nu_a=0$.
Thus the final number of zero modes is further reduced by half
after rotation. For example, in the cases of
$D3$-brane with $\Omega_0 = \Omega_\pi$ and $D7$-brane with
$\Omega_0 = \gamma^{245678}$, there are four zero modes
while, in the cases of $D3$-brane with $\Omega_0 = \pm \gamma^{24}$, $D5$-brane with
$\Omega_0 = \pm \gamma^{2345}$, and $D7$-brane with
$\Omega_0 = \pm \gamma^{135678}$, there exist two zero modes.

Now we consider $D_+ p - (+,-,2,2)$ brane intersections. We assume
$(+,-,2,2)$-brane is oriented along the $X^{1,3}$-axes and
$X^{5,7}$-axes. In this case there are two possibilities rotating
the $(+,-,2,2)$-brane while preserving supersymmetry. One
is an $SU(2)$ rotation as in Eq. \eq{R}. The other is an $SU(2) \times SU(2)$
rotation described by
\be \la{double-R}
R = R_1 R_2 = e^{\half(\gamma^{12} \phi_1 + \gamma^{34} \phi_2)}
e^{\half(\gamma^{56} \phi_3 + \gamma^{78} \phi_4)}
\ee
whose eigenvalues are
\be \la{rep-R-22}
R = e^{i\sum_{i=1}^4 s_i \phi_i}.
\ee
Since the single $SU(2)$ rotation is almost the same as the
previous $SU(2)$ case, we will analyze the double $SU(2)$ case
only. The mode expansion of complex
bosonic coordinates $Z^i \;(i=1,\cdots,4)$ where $Z^3 = X^5 + i X^6$ and
$Z^4 = X^7 +i X^8$ has the same form as Eqs. \eq{mode-complex}
and \eq{dn-mode-complex} with the angles
$0 \leq \delta_i = \phi_i/\pi \leq \half$.

Since $R$ is a trivial rotation in the case $[R, \Omega_0]=0$,
we restrict attention to the case satisfying the following
condition
\be \la{ro-gamma-omega-22}
\{\gamma^{12}, \Omega_0 \} = \{ \gamma^{34}, \Omega_0 \} =
\{\gamma^{56}, \Omega_0 \} = \{ \gamma^{78}, \Omega_0 \} = 0.
\ee
Only the $D_+ 5 =(+,-,2,2)$-brane belongs to
the nontrivial case of Eq. \eq{ro-gamma-omega-22}.
Note that $(+,-,1,1)$- and $(+,-,3,3)$-branes are excluded even in
the single $SU(2)$ rotation
since a nontrivial rotation which is $U(1)$ or $U(1) \times U(1)$
completely breaks the supersymmetry.
The mode expansion is given by Eq. \eq{spinor-d+} with
\bea \la{ro-fermion-22}
&& \xi^1 (\tau, \sigma)= \xi^1_0 (\tau, \sigma)
+ \sum_{\k}c_\k(\varphi_\k^1(\tau,\sigma) \Omega_0
\widetilde{S}_\k
+i \rho_\k \varphi_\k^2(\tau,\sigma)\Pi S_\k), \xx
&& \xi^2 (\tau, \sigma)= \xi^2_0 (\tau, \sigma)
+ \sum_{\k}c_\k(\varphi_\k^2(\tau,\sigma) S_\k
- i \rho_\k \varphi_\k^1(\tau,\sigma)\Pi \Omega_0 \widetilde{S}_\k), \xx
&& \eta^1 (\tau, \sigma)= \sum_{\lambda}c_\lambda( \varphi_\lambda^1(\tau,\sigma)
\Omega_0 \widetilde {S}_\lambda
+ i \rho_\lambda \varphi_\lambda^2(\tau,\sigma)\Pi S_\lambda ), \xx
&& \eta^2 (\tau, \sigma)= \sum_{\lambda}c_\lambda (\varphi_\lambda^2(\tau,\sigma)
S_\lambda - i \rho_\lambda \varphi_\lambda^1(\tau,\sigma)\Pi
\Omega_0 \widetilde{S}_\lambda ),
\eea
where the mode numbers $\k$ and $\lambda$ are now determined
by the following equations
\bea \la{klam-22}
&& e^{2\pi i \k} = \frac{e^{2\pi i \nu_a} \mp i \rho_\k}
{1 \pm i e^{2\pi i \nu_a}\rho_\k}
\frac{\frac{\k}{|\a|} \pm i \mu}{\omega_\k},
\qquad \;\; \mbox{for} \; S_\k^\pm, \\
&& e^{2\pi i \lambda} = - \frac{e^{2\pi i \nu_a} \pm i \rho_\lambda}
{1 \mp i e^{2\pi i \nu_a}\rho_\lambda}
\frac{\frac{\lambda}{|\a|} \pm i \mu}{\omega_\lambda},
\qquad \mbox{for} \; S_\lambda^\pm
\eea
with $\nu_a = \sum_{i=1}^4 s_i \delta_i$.
We decomposed the spinors $S_\k$ and $S_\lambda$ as in Eq.
\eq{dec-d+}:
\be \la{dec-22}
S_\k^\pm = \half(1 \pm \Pi\Omega_0) S_\k,
\qquad S_\lambda^\pm = \half(1 \pm \Pi\Omega_0) S_\lambda.
\ee

If $\nu_a=0$, one can see that there can be
zero modes since $\k =0$ is a solution of Eq. \eq{klam-22}. They are
of the same form as the zero modes in Eq. \eq{spinor-d+}, but with
$S_0$ satisfying $R^2 S_0 = S_0$. Thus the final number of zero
modes is further reduced by quarter after the $SU(2) \times SU(2)$ rotation,
which is the same situation as the flat spacetime \ct{bdl,polchinski}.
For example, in the cases with $\sharp_{ND}= 0,8$, namely, $\Omega_0 = \Omega_\pi$
and $\Omega_0 = \Omega_\pi \gamma$, there are two zero modes
while, in the case with $\sharp_{ND} = 4$,
there exist only one zero mode.

%%%%%%%%%%%%%%%%%%%%%%%%%%%%%%%%%%%%%%%%%%%%%%%%%%%%%%%%%%%%%%%%%%%%%%
\section{Supersymmetry of Intersecting D-branes}
%%%%%%%%%%%%%%%%%%%%%%%%%%%%%%%%%%%%%%%%%%%%%%%%%%%%%%%%%%%%%%%%%%%%%%

In a light-cone gauge, the 32 components of the supersymmetries
for closed strings decompose into kinematical supercharges,
$Q^{+A}_a$, and dynamical supercharges, $Q^{-A}_{\ad}$.
For superstrings in the plane wave background with
the action \eq{gs-action}, the conserved super-N\"other
charges were identified by Metsaev \ct{metsaev1}:
\bea \la{charge-p}
&& Q^{+1} = \frac{\sqrt{2p^+}}{2\pi \a^\prime p^+}
\int_{0}^{2\pi \a^\prime |p^+|} d\sigma (\cos\mu\tau S^1 -\sin \mu\tau \Pi S^2), \\
\la{charge-q+2}
&& Q^{+2} = \frac{\sqrt{2p^+}}{2\pi \a^\prime p^+}
\int_{0}^{2\pi \a^\prime |p^+|} d\sigma (\cos\mu\tau S^2 + \sin \mu\tau \Pi S^1), \\
\la{charge-q-1}
&& \sqrt{2p^+}Q^{-1}= \frac{1}{2\pi \a^\prime p^+}
\int_{0}^{2\pi \a^\prime |p^+|} d\sigma \Bigl( \p_- X^I\gamma^I S^1
-\mu X^I\gamma^I \Pi S^2 \Bigr), \\
\la{charge-q-2}
&& \sqrt{2p^+}Q^{-2}= \frac{1}{2\pi \a^\prime p^+}
\int_{0}^{2\pi \a^\prime |p^+|} d\sigma \Bigl( \p_+ X^I\gamma^I
S^2 + \mu X^I\gamma^I \Pi S^1 \Bigr).
\eea

The super-N\"other charges of an open string are given by a
subset of the symmetries of the closed string action which are
compatible with the open string boundary conditions.
Due to the boundary condition \eq{bc-fermion},
it turns out that the conserved dynamical supercharge is given by
the combination
\bea \la{dyn-susy}
q^- &=& \frac{1}{\pi |\a|} \int^{\pi |\a|}_0 d \sigma q_\tau^- \xx
&=& I_+(Q^{-2}- \Omega_\theta^T Q^{-1}).
\eea
It was shown that all half BPS D-branes in the type IIB plane wave
background have to satisfy
\be \la{bc-}
X^{r^\prime}|_{\p \Sigma}=0, \qquad \forall \; {r^\prime}
\ee
for the Dirichlet coordinates of $D_-$-branes \ct{skenderis1}, and
\be \la{bc+}
(\p_\sigma X^r \Omega^T_\theta S^1- \mu X^r \Pi S^1 )|_{\p
\Sigma}=0, \qquad \forall \; r
\ee
for the Neumann coordinates of $D_+$-branes \ct{kim}.

Although $D_-$-branes located at a constant
transverse position $x^{r^\prime}_0 \neq 0$ superficially appear to break
all dynamical supersymmetries, the broken dynamical
supersymmetries can be restored by
incorporating a worldsheet symmetry \ct{skenderis1}.
The superstring action \eq{gs-action}
is invariant under an arbitrary shift of the field by a parameter
that satisfies the same field equation and open string boundary
condition as the original field. Using this fact, one can find
modified transformation rules by using the worldsheet symmetry,
which now lead to a conserved charge. On the other hand, one
cannot use the worldsheet symmetry to restore some apparently
broken dynamical supersymmetry for $D_+$-branes since
the symmetry breaking terms involve Neumann coordinates as shown in Eq. \eq{bc+}.
Only special classes of $D_+$-branes allow the condition \eq{bc+}.
These are $D1$-branes in which, by definition, $X^r=0$ for all $r$
and $D5$-branes of type $(+,-,4,0)$ or $(+,-,0,4)$, thus $\Omega^T_\theta S^1 = \Pi S^1$,
with a Born-Infeld flux satisfying
the modified Neumann boundary condition \eq{+d5-nbc}.
Another $D_+$-branes cannot satisfy the condition \eq{bc+}
and thus the dynamical supersymmetry is not conserved.

Keeping these facts in mind, one may deduce some conditions for
supersymmetric intersection of D-branes. The dynamical
supersymmetry of intersecting $D_- - D_-$ branes can be preserved
only when they have coincident transverse positions, since the
worldsheet shift symmetry must be simultaneously applied to both
branes. Even parallel but separated $D_-$-branes of the same type
preserve no dynamical supersymmetry. However, the dynamical
supersymmetry of intersecting $D_+$-branes does not depend on
their transverse locations.

We now give a rigorous worldsheet derivation on conserved
supercharges for each case of intersecting D-branes. The conserved
kinematical supersymmetry depends on the type of
D-branes as shown in \ct{skenderis1,kim}. Let us first
consider $D_- - D_-$ brane configurations. The conserved
kinematical supercharge is given by the form
\bea \la{kin-susy}
q^+ &=& \frac{1}{\pi |\a|} \int^{\pi |\a|}_0 d \sigma q_\tau^+ \xx
&=& \left\{%
\begin{array}{ll}
    I_+(Q^{+1} + \Omega_\theta Q^{+2}), & \hbox{for A-type,} \\
    I_-(Q^{+1} + \Omega_\theta Q^{+2}), & \hbox{for B-type.} \\
\end{array}%
\right.
\eea
Using the equations of motion, Eqs. \eq{eom-boson} and \eq{eom-fermion},
it is easy to show that the kinematical supercharge density $q^+_\tau$
in Eq. \eq{kin-susy} satisfies the following conservation law
\be \la{con-law-kin}
\frac{\p q^+_\tau}{\p \tau} + \frac{\p q^+_\sigma}
{\p \sigma} =0,
\ee
with
\be \la{current-kin}
q^+_\sigma = \sqrt{2p^+}
\Bigl( e^{\mu \tau \Omega_\theta \Pi}I_{\pm}(S^1 - \Omega_\theta S^2)
\Bigr),
\ee
where we used the fact that $I_{\pm}$ commute with $\Omega_\theta \Pi$.
Then one can see that $q^+$ is conserved by observing that $q^+_\sigma$ vanishes
at boundaries for A-type branes with $I_+$ and for B-type branes
with $I_-$. Now one can easily count the remaining kinematical
supersymmetries for each case in \eq{omega-sol} and
the components of conserved supersymmetries can be easily
identified using the projection matrices $I_\pm$. For $\Omega_0 = \Omega_\pi$ and
$\Omega_0 = \Omega_\pi \gamma$, $I_+$ becomes identity and
we have 8 kinematical supersymmetries. For $\Omega_0 = - \Omega_\pi$
and $\Omega_0 = - \Omega_\pi \gamma$, $I_+$ becomes identically zero and so no kinematical
supersymmetry is preserved. In the case of $\Omega_0 = \pm \Omega_\pi \Xi$, we have 4
kinematical supersymmetries since $I_+ =\half(I_8 \pm \Xi)$. The conserved kinematical
supersymmetry of intersecting $D_-$-branes is independent of their
transverse locations.

It is useful to recall the (anti-)commutation relations between
$\gamma^I = \{\gamma^r, \gamma^{r^\prime}, \gamma^i,
\gamma^{i^\prime} \}, \; \Omega_0$ and $\Omega_\pi$
to find conserved dynamical supersymmetries:
\bea \la{gamma-omega-comm}
&& \{\gamma^r, \Omega_0 \}= \{\gamma^i, \Omega_0 \}
=[\gamma^{r^\prime}, \Omega_0 ]= [ \gamma^{i^\prime}, \Omega_0 ]=0, \\
\la{gamma-inik}
&& \{\gamma^r, \Omega_\pi \}= \{\gamma^{i^\prime}, \Omega_\pi \}
=[\gamma^{r^\prime}, \Omega_\pi ]= [ \gamma^i, \Omega_\pi ]=0.
\eea
Using the similar recipe used in the above kinematical
supersymmetry, it is not difficult to show that the dynamical supercharge density $q^-_\tau$
in Eq. \eq{dyn-susy} also satisfies the conservation law
\be \la{con-law-dyn}
\frac{\p q^-_\tau}{\p \tau} + \frac{\p q^-_\sigma}
{\p \sigma} = 0,
\ee
where
\bea \la{ds-current}
q^-_\sigma &=& \sqrt{\frac{1}{2p^+}}
\Bigl( (\p_\tau X^r \gamma^r \Omega^T_\theta + \mu X^r \gamma^r \Pi)
I_{+,-}(S^1 - \Omega_\theta S^2)
 -\p_\sigma X^r \gamma^r \Omega^T_\theta I_{+,-}(S^1 + \Omega_\theta S^2) \xx
&& -(\p_\tau X^{r^\prime} \gamma^{r^\prime} \Omega^T_\theta
- \mu X^{r^\prime} \gamma^{r^\prime} \Pi)
I_{+,-}(S^1 + \Omega_\theta S^2)
 + \p_\sigma X^{r^\prime} \gamma^{r^\prime} \Omega^T_\theta I_{+,-}(S^1 -
 \Omega_\theta S^2) \\
&& + e^{i \theta}(\p_\tau X^i \gamma^i \Omega^T_\theta + e^{i \theta}\mu X^i \gamma^i \Pi)
I_{-,+}(S^1 - e^{i \theta}\Omega_\theta S^2)
 -e^{i \theta}\p_\sigma X^i \gamma^i \Omega^T_\theta I_{-,+}
 (S^1 + e^{i \theta}\Omega_\theta S^2) \xx
&& -e^{i \theta}(\p_\tau X^{i^\prime} \gamma^{i^\prime} \Omega^T_\theta
- e^{i \theta}\mu X^{i^\prime} \gamma^{i^\prime} \Pi)
I_{-,+}(S^1 + e^{i \theta}\Omega_\theta S^2)
 + e^{i \theta}\p_\sigma X^{i^\prime} \gamma^{i^\prime}
 \Omega^T_\theta I_{-,+}(S^1 - e^{i \theta}\Omega_\theta
 S^2) \Bigr) \nonumber
\eea
with the first subscript in $I_{+,-}$ or $I_{-,+}$ for A-type
branes while with the second one for B-type branes.
Here we presented the general expression for $\Omega_\theta =(\Omega_0, \Omega_\pi)$
for later use. Since one can modify the transformation rules in a
way to recover the dynamical supersymmetry for $D_-$-branes
located off origin, as discussed before, here we assume without
loss of generality that the two branes are placed at origin, viz.,
\be \la{at-origin}
X^{r^\prime}|_{\p \Sigma} = X^{i}|_{\sigma=\pi \a}=
X^{i^\prime}|_{\sigma=0} = 0.
\ee
Using the crucial identities \eq{identity-a} and \eq{identity-b}
as well as the open string boundary conditions,
\eq{nbc-boson}-\eq{bc-fermion}, and \eq{at-origin},
it is easy to see that
\be \la{d-current=0}
q^-_\sigma|_{\p \Sigma} = 0.
\ee

Now it is simple to identify the
conserved dynamical supersymmetry for each case in Eq. \eq{omega-sol}.
Obviously the $\Omega_0 = \Omega_\pi$ case preserves
8 dynamical supersymmetries whereas the $\Omega_0 = - \Omega_\pi$ case
does not preserve any dynamical supersymmetry as expected. This
should be the case since the former corresponds to a coincident
$Dp$-$Dp$ brane configuration while the latter does to a
$Dp$-anti-$Dp$
brane configuration. Similarly, when $\Omega_0 = - \Omega_\pi
\gamma$, 8 dynamical supersymmetries are presereved while the
dynamical supersymmetries are completely broken in the case of
$\Omega_0 = \Omega_\pi \gamma$.
Finally, the cases of $\Omega_0^T \Omega_\pi = \pm \Xi$
preserve 4 dynamical supersymmetries. The results are summarized
in Table 1.

We next consider $D_+ - D_+$ brane configurations. The kinematical
supercharge density
\be \la{kin-susy-d+}
q^+_\tau = \sqrt{2p^+} \sqrt{\frac{\pi \mu |\a|}{\sinh \pi \mu |\a|}} \,
e^{\mu(\sigma-\half \pi|\a|) \Omega_\theta \Pi}I_\pm(S^1 + \Omega_\theta S^2),
\ee
satisfies the conservation law \ct{kim}
\be \la{conser-d+}
\frac{\p q^+_\tau}{\p \tau} + \frac{\p q^+_\sigma}
{\p \sigma}=0
\ee
with
\be \la{kin-current-d+}
q^+_\sigma = \sqrt{2p^+} \sqrt{\frac{\pi \mu |\a|}{\sinh \pi \mu |\a|}} \,
e^{\mu(\sigma-\half \pi|\a|) \Omega_\theta \Pi}I_\pm(S^1 - \Omega_\theta S^2).
\ee
The corresponding kinematical supersymmetry can be easily seen to
be conserved with $I_+$ for A-type branes and $I_-$ for B-type
branes, using Eqs. \eq{identity-a} and \eq{identity-b}.
One can see that the cases $\Omega_0 = \Omega_\pi$
and $\Omega_0 = \Omega_\pi \gamma $ preserve the 8 kinematical
supersymmetries irrespective of their transverse locations
while 4 supersymmetries for the cases $\Omega_0 = \pm \Omega_\pi \Xi$.
The other cases totally break the kinematical supersymmetry. This
is the same as those in $D_- - D_-$.

To find the conserved dynamical supersymmetry one can apply the
same procedure as in the $D_-$-brane case. As already mentioned,
only $D1$-branes and $D5$-branes with flux can preserve the
dynamical supersymmtry. Hence it is sufficient to consider A-type
branes only. After some calculation, one gets the result:
\be \la{con-law-dyn-d+d+}
\frac{\p q^-_\tau}{\p \tau} + \frac{\p q^-_\sigma}
{\p \sigma} = 0,
\ee
where
\bea \la{ds-current-d+}
q^-_\sigma &=& \sqrt{\frac{1}{2p^+}}
\Bigl( -(\p_\sigma X^r \gamma^r \Omega^T_\theta - \mu X^r \gamma^r \Pi)
I_{+}(S^1 + \Omega_\theta S^2)
 +\p_\tau X^r \gamma^r \Omega^T_\theta I_{+}(S^1 - \Omega_\theta S^2) \xx
&& + (\p_\sigma X^{r^\prime} \gamma^{r^\prime} \Omega^T_\theta
+ \mu X^{r^\prime} \gamma^{r^\prime} \Pi)
I_{+}(S^1 - \Omega_\theta S^2)
 - \p_\tau X^{r^\prime} \gamma^{r^\prime} \Omega^T_\theta
 I_{+}(S^1 + \Omega_\theta S^2) \\
&& - e^{i\theta}(\p_\sigma X^i \gamma^i \Omega^T_\theta -
e^{i \theta} \mu X^i \gamma^i \Pi)
I_{-}(S^1 + e^{i \theta}\Omega_\theta S^2)
 + e^{i\theta} \p_\tau X^i \gamma^i \Omega^T_\theta I_{-}
 (S^1 -e^{i\theta} \Omega_\theta S^2) \xx
&& + e^{i\theta} (\p_\sigma X^{i^\prime} \gamma^{i^\prime} \Omega^T_\theta
+ e^{i\theta} \mu X^{i^\prime} \gamma^{i^\prime} \Pi)
I_{-}(S^1 - e^{i \theta} \Omega_\theta S^2)
 - e^{i\theta} \p_\tau X^{i^\prime} \gamma^{i^\prime}
 \Omega^T_\theta I_{-}(S^1 +e^{i\theta}
 \Omega_\theta S^2) \Bigr) \nonumber.
\eea
In the case of $\Omega_0 = \Omega_\pi$, that is, parallel $D1$-$D1$ and $D5$-$D5$ branes,
one can see that
$q^-_\sigma|_{\p \Sigma} = 0$, so 8 dynamical supersymmetries are
preserved.
In the case of $\Omega_0^T \Omega_\pi= \pm
\Xi$, however, $q^-_\sigma$ at the boundary $\p \Sigma$ reduces to
\be \la{bd-d5d1}
q^-_\sigma|_{\p \Sigma} =\sqrt{\frac{1}{2p^+}}
(\p_\sigma X^{i^\prime} \gamma^{i^\prime} \Omega^T_0
+  \mu X^{i^\prime} \gamma^{i^\prime} \Pi)
I_-(S^1 - \Omega_0 S^2)
\ee
for $\Omega_\theta = \Omega_0$, while, for $\Omega_\theta = \Omega_\pi$,
\be \la{bd-d1d5}
q^-_\sigma|_{\p \Sigma} =  \sqrt{\frac{1}{2p^+}}
(\p_\sigma X^i \gamma^i \Omega^T_\pi + \mu X^i \gamma^i \Pi)
I_-(S^1 - \Omega_\pi S^2).
\ee
Therefore one can find the following results for the dynamical
supersymmetries of $D1$-$D5$ branes:
\bea \la{dyn-susy-d+d+}
&& q^-= I_+(Q^{-2} - \Omega_0^T Q^{-1}), \qquad \mbox{for} \;
D5-D1, \xx
&& q^-= I_+(Q^{-2} - \Omega_\pi^T Q^{-1}), \qquad \mbox{for} \;
D1-D5,
\eea
where only 4 dynamical supersymmtries are conserved for the
two cases since $I_+ = \half(1 \pm \Xi)$.

When $\Omega_0 = - \Omega_\pi \gamma$ corresponding to $(+,-,4,0)-(+,-,0,4)$ brane configurations,
only the last two lines in Eq. \eq{ds-current-d+} are relevant and we meet an
intriguing situation. Some parts in Eq.
\eq{ds-current-d+} (explicitly, \eq{bd-d5d1} for $\Omega_\theta =
\Omega_0$ and \eq{bd-d1d5} for $\Omega_\theta =
\Omega_\pi$) no longer vanish at the boundary $\p \Sigma$.
The dynamical supersymmetry of $(+,-,4,0)-(+,-,0,4)$ brane
intersections is thus completely broken. This is certainly caused
by worldvolume fluxes in $D5$-branes. The results are summarized
in Table 2.

In section 3 we analyzed D-branes intersecting at general angles.
The supersymmetric intersecting D-branes at general angles are
possible only for specific branes such as $D3$- and $D7$-branes
among $D_-$-branes and $(+,-,2,2)$-branes among $D_+$-branes. It
turned out that under $SU(2)$ rotations the half of spinors have
the same mode expansion as before the rotation
while under $SU(2) \times SU(2)$ rotations the quarter of
spinors have the same mode expansion. Thus we naturally expect
that the conserved supersymmetry is further reduced by half in the
case of single $SU(2)$ rotation and by
quarter in the case of double $SU(2)$ rotation \ct{bdl,polchinski,smith}.
This should be the case since the analysis about conserved
supersymmetries for intersecting D-branes at general angles is
identical to parallel D-branes or D-branes intersecting at right
angles except that the
boundary condition at $\sigma = \pi |\a|$ is instead given
by the second equation in \eq{bc-ro-fermion} and thus a further constraint
$(R^2 S^2 - S^2)_{\sigma = \pi |\a|} = 0$ is needed to satisfy
the boundary condition at $\sigma = \pi |\a|$. This additional condition reduces the
number of spinors satisfying $q_\sigma^\pm|_{\p \Sigma} = 0$
by half in the case of single $SU(2)$ rotation while by quarter in
the case of double $SU(2)$ rotation as mentioned above. This is
the same situation encountered in flat spacetime. We summarized
our results for the remaining supersymmetries of all possible
supersymmetric intersecting D-branes in Tables 1-2.

\begin{table}[tbp]
\begin{center}
\begin{tabular}{|c|c|c|c|c|} \hline
\multicolumn{1}{|c|}{} &
\multicolumn{1}{|c|}{$\sharp_{ND}$} &
\multicolumn{1}{|c|}{$\Omega_0^T \Omega_\pi$} &
\multicolumn{1}{|c|}{$q^+$} &
\multicolumn{1}{|c|}{$q^-$} \\ \hline
& 0 & 1 & 8 & 0 \\ \cline{3-5}
& & $-1$  & 0  & 0 \\ \cline{2-5}
& 4 & $\Xi$ & 4 & 0 \\ \cline{3-5}
$Dp \leftrightarrow Dq$ & & $-\Xi$ & 4 & 0 \\ \cline{2-5}
& 8 & $\gamma$ & 8 & 0 \\ \cline{3-5}
&  & $-\gamma$ & 0 & 0 \\ \hline
& 0 & 1 & 4 & 0 \\ \cline{3-5}
& & $-1$  & 0  & 0 \\ \cline{2-5}
$Dp \leftrightarrow D^\prime q $ & 4 & $\Xi$ & 2 & 0 \\ \cline{3-5}
$ p,q = 3,7 $ & & $-\Xi$ & 2 & 0 \\ \cline{2-5}
& 8 & $\gamma$ & 4 & 0 \\ \cline{3-5}
&  & $-\gamma$ & 0 & 0 \\ \hline
& 0 & 1 & 8 & 8 \\ \cline{3-5}
& & $-1$  & 0  & 0 \\ \cline{2-5}
& 4 & $\Xi$ & 4 & 4 \\ \cline{3-5}
$Dp - Dq$ & & $-\Xi$ & 4 & 4 \\ \cline{2-5}
& 8 & $\gamma$ & 8 & 0 \\ \cline{3-5}
&  & $-\gamma$ & 0 & 8 \\ \hline
& 0 & 1 & 4 & 4 \\ \cline{3-5}
& & $-1$  & 0  & 0 \\ \cline{2-5}
$Dp - D^\prime q$ & 4 & $\Xi$ & 2 & 2 \\ \cline{3-5}
$ p,q = 3,7 $ & & $-\Xi$ & 2 & 2 \\ \cline{2-5}
& 8 & $\gamma$ & 4 & 0 \\ \cline{3-5}
&  & $-\gamma$ & 0 & 4 \\ \hline
\end{tabular}
\end{center}
\caption{Supersymmetry in $D_- - D_-$ intersections. Separated and coincident
D-branes are denoted by the bi-arrow and the dash, respectively. A
rotated D-brane by $SU(2)$ is indicated by the prime.}
\label{tableone}
\end{table}
\begin{table}[tbp]
\begin{center}
\begin{tabular}{|c|c|c|c|c|} \hline
\multicolumn{1}{|c|}{} &
\multicolumn{1}{|c|}{$\sharp_{ND}$} &
\multicolumn{1}{|c|}{$\Omega_0^T \Omega_\pi$} &
\multicolumn{1}{|c|}{$q^+$} &
\multicolumn{1}{|c|}{$q^-$} \\ \hline
& 0 & 1 & 8 & 0 \\ \cline{3-5}
& & $-1$  & 0  & 0 \\ \cline{2-5}
$Dp \leftrightarrow Dq$ & 4 & $\Xi$ & 4 & 0 \\ \cline{3-5}
$p \neq 1, q \neq 1$ & & $-\Xi$ & 4 & 0 \\ \cline{2-5}
& 8 & $\gamma$ & 8 & 0 \\ \cline{3-5}
&  & $-\gamma$ & 0 & 0 \\ \hline
& 0 & 1 & 4 & 0 \\ \cline{3-5}
& & $-1$  & 0  & 0 \\ \cline{2-5}
$(+,-,2,2)\leftrightarrow (+,-,2,2)^\prime $ & 4 & $\Xi$ & 2 & 0 \\ \cline{3-5}
 &  & $-\Xi$ & 2 & 0 \\ \cline{2-5}
& 8 & $\gamma$ & 4 & 0 \\ \cline{3-5}
&  & $-\gamma$ & 0 & 0 \\ \hline
& 0 & 1 & 2 & 0 \\ \cline{3-5}
& & $-1$  & 0  & 0 \\ \cline{2-5}
$(+,-,2,2) \leftrightarrow (+,-,2,2)^{\prime \prime} $ & 4 & $\Xi$ & 1 & 0 \\ \cline{3-5}
 &  & $-\Xi$ & 1 & 0 \\ \cline{2-5}
& 8 & $\gamma$ & 2 & 0 \\ \cline{3-5}
&  & $-\gamma$ & 0 & 0 \\ \hline
$D1 \leftrightarrow D1$ & 0 & 1 & 8 & 8 \\ \cline{3-5}
& & $-1$  & 0  & 0 \\ \hline
$D1 \leftrightarrow \widetilde{D5}$ & 4 & $\Xi$ & 4 & 4 \\ \cline{3-5}
& & $-\Xi$ & 4 & 4 \\ \hline
& 0 & 1  & 8  & 8 \\ \cline{3-5}
$\widetilde{D5}\leftrightarrow \widetilde{D5}$ &  & $-1$ & 0 & 0 \\ \cline{2-5}
& 8 & $\gamma$ & 8 & 0 \\ \cline{3-5}
&  & $-\gamma$ & 0 & 0 \\ \hline
\end{tabular}
\end{center}
\caption{Supersymmetry in $D_+ - D_+$ intersections. We used the
bi-arrow to emphasize that the supersymmetry is independent of
tranverse locations. $(+,-,2,2)^{\prime}$ is rotated by $SU(2)$ while
$(+,-,2,2)^{\prime \prime}$ is rotated by $SU(2) \times SU(2)$.
A $D5$-brane with flux is indicated by the widetilde.}
\label{tabletwo}
\end{table}

In the plane wave background \eq{pp-metric}, $D_-$9-brane is not supersymmetric
and $D_+$9-brane is at most a quarter BPS. Thus it is not trivial
to say about type I superstring theory in the plane wave
background since it is not obvious how to introduce $D9$-branes
carrying $SO(32)$ Chan-Paton factors. Furthermore T-dualty
transformation in the plane wave background is quite different
from flat spacetime case since there is a RR 4-form and the
geometry is curved. (It should be interesting to clarify these
open problems in the type IIB plane wave background.)
Nevertheless, we would like to use the flat spacetime analogue to
plainly explain our results in Tables 1-2. The type I string
theory contains 16 $D9$-branes carrying
$SO(32)$ Chan-Paton factors, and $D1$-brane ($\Omega_0^T \Omega_\pi = \gamma$)
and anti-$D1$-brane ($\Omega_0^T \Omega_\pi = -\gamma$) are both half-BPS
states, preserving 8 supersymmetries \ct{bachas}.
Of course, if $D1$-brane and anti-$D1$-brane coexist in type I or
type IIB string theory, they totally break the supersymmetry.
Similarly,
$D5$-$D1$ ($\Omega_0^T \Omega_\pi = \Xi$)
corresponds to instantons while $D5$-anti-$D1$ ($\Omega_0^T \Omega_\pi = -\Xi$)
does to anti-instantons in type IIB string theory, but both
instantons and anti-instantons are BPS states preserving 8
supersymmetries in Yang-Mills gauge theory \ct{bvs}. And other
supersymmetric intersecting D-branes with $\sharp_{ND} =4,8$ in Tables
1-2 are plane wave analogues of the D-branes generated by
T-duality from the previous examples in flat spacetime.

%%%%%%%%%%%%%%%%%%%%%%%%%%%%%%%%%%%%%%%%%%%%%%%%%%%%%%%%%%%%%%%%%%%%%%
\section{Supersymmetry Algebra of Intersecting D-branes}
%%%%%%%%%%%%%%%%%%%%%%%%%%%%%%%%%%%%%%%%%%%%%%%%%%%%%%%%%%%%%%%%%%%%%%

In this section we will present the explicit mode expansions and
their supersymmetry algebra of conserved supercharges $q^\pm$ only
for parallel and orthogonally intersecting D-branes for
simplicity. The generalization to intersecting D-branes at general
angles should be straightforward.

Let us explain how to easily calculate the mode expansion of
dynamical supercharge $q^-$ without a little tedious manipulation.
In section 2 we determined the mode expansions of an open string
stretched between $Dp$-brane and $Dq$-brane to satisfy the
equations of motion and boundary conditions. And it was shown in
section 4 that dynamical supercharges satisfy conservation
law such as Eqs. \eq{con-law-dyn}, \eq{con-law-dyn-d+d+},
and the spatial worldsheet current $q_\sigma^-$
vanishes at worldsheet boundaries.
This immediately implies that the charge $q^-$ in Eq. \eq{dyn-susy}
is conserved, in other words, it should be $\tau$-independent. From the
open string mode expansion, one can see that the time dependent
factor is always of the form $e^{-i(\omega_\k + \omega_\lambda) \tau}$ where
$\omega_\k$ is from, say, a bosonic mode and $\omega_\lambda$ is from
a fermionic mode. If $\k + \lambda \neq 0$, we have $\tau$-dependent terms and
thus these terms are necessarily cancelled with another terms.

%%%%%%%%%%%%%%%%%%%%%%%%%%%%%%%%%%%%%%%%%%%%%%%%%%%%%%%%%%%%%%%%%%%%%%
\subsection{$D_- - D_-$}
%%%%%%%%%%%%%%%%%%%%%%%%%%%%%%%%%%%%%%%%%%%%%%%%%%%%%%%%%%%%%%%%%%%%%%

From the previous analysis, we have learned that $D_- - D_-$ brane
configurations are mostly close to the flat spacetime case. The
kinematical and the dynamical supercharges in this case are given
by
\bea \la{mode-kin-d-d-}
&& q^+ = 2 \sqrt{2p^+} I_\pm S_0, \xx
&& \sqrt{2p^+} q^- = 2I_+ \Bigl( p_0^r \gamma^r \Omega_0^T S_0
+ \mu x_0^r \gamma^r \Pi S_0
+ \sum_{\k} c_\k \a_{-\k}^I \gamma^I S_{\k}
- \frac{i \mu}{2c_\k \omega_\k}\a_{-\k}^I \Omega_0^T \gamma^I \Pi S_{\k}
\Bigr),
\eea
where the index $I$ collectively denotes NN, ND, DN and DD
indices; $I=\{ r, i, i^\prime, r^\prime \}$ and
the matrix $I_+$ in $q^+$ is for A-type branes while $I_-$ for
B-type branes. Here the mode number
$\k$ should be understood as nonzero integers for $I=\{ r, r^\prime \}$
and half-integers for $I=\{i, i^\prime \}$.
Then the supersymmetry algebra reads as
\bea \la{alg-kin-d-d-}
&& \{q^+_a, q^+_b \} = 2p^+ (I_\pm)_{ab}, \\
&& \{q^+_a, q^-_{\ad}\}= (I_+ \Omega_0 \gamma^r)_{a\ad}  P^r - \frac{\mu}{p^+}
(I_+ \Pi \gamma^r)_{a\ad} J^{+r}, \\
\la{dyn-qq}
&& \{q^-_{\ad}, q^-_{\bd}\}= 2H (I_+)_{\ad\bd} + \frac{\mu}{2p^+}
\Bigl( (I_+ \gamma^{JK}_I \Pi\Omega_0)_{\ad\bd} J^{JK}_I -
(I_+ \gamma^{JK}_{II} \Pi\Omega_0)_{\ad\bd} J^{JK}_{II} \Bigr),
\eea
where $P^r = p_0^r$ and $J^{+r}=-x_0^r p^+$ are translational and
boost generators along NN directions, respectively. The
hamiltonian $H$ is given by
\bea \la{d-d-H}
2p^+ H &=& \half(p_{0r}^2 + \mu^2 x_{0r}^2) - 2\mu i S_0 \Omega_0 \Pi I_\pm S_0
+ \sum_{n \neq 0} \Bigl( \sum_{I=\{ r, r^\prime \}}
\half \a_{-n}^I \a_n^I + 2 \omega_n S_{-n} I_+ S_n \Bigr) \xx
&& + \sum_{\k} \Bigl( \sum_{I=\{ i, i^\prime \}} \half \a_{-\k}^I \a_\k^I
+ 2 \omega_\k S_{-\k} I_- S_\k \Bigr)
\eea
and $J^{JK}_I$ and $J^{JK}_{II}$ are rotational generators in NN,
ND, DN and DD directions in the first $SO(4)$ directions
and the second $SO(4)^\prime$ directions, respectively,
whose expressions in terms of modes are given by
\bea \la{d-d-Jn}
J^{JK} &=& x_0^r p_0^s - x_0^s p_0^r - i S_0 \gamma^{JK}I_\pm S_0
- i \sum_{\k} S_{-\k} \gamma^{JK} I_- S_\k \xx
&&- i \sum_{n \neq 0} \Bigl( \frac{1}{2\omega_n}
(\a_{-n}^J \a_n^K - \a_{-n}^K \a_n^J)
+  S_{-n} \gamma^{JK} I_+ S_n \Bigr)
\eea
for $J^{JK} = \{ J^{rs}, J^{r^\prime s^\prime}\}$ and
\bea \la{d-d-Jk}
J^{JK} &=& - i S_0 \gamma^{JK}I_\pm S_0
- i \sum_{n \neq 0} S_{-n} \gamma^{JK} I_+ S_n \xx
&& - i \sum_{\k} \Bigl( \frac{1}{2\omega_\k}
(\a_{-\k}^J \a_\k^K - \a_{-\k}^K \a_\k^J)
+  S_{-\k} \gamma^{JK} I_- S_\k \Bigr)
\eea
for $J^{JK} = \{ J^{ij}, J^{i^\prime j^\prime}\}$.
In Appendix A, we illustrate the superalgebra \eq{dyn-qq} for a specific example, $D3-D5$,
since some nontrivial identities have been used to derive it.

%%%%%%%%%%%%%%%%%%%%%%%%%%%%%%%%%%%%%%%%%%%%%%%%%%%%%%%%%%%%%%%%%%%%%%
\subsection{$D_+ - D_+$}
%%%%%%%%%%%%%%%%%%%%%%%%%%%%%%%%%%%%%%%%%%%%%%%%%%%%%%%%%%%%%%%%%%%%%%

We concentrate only on the $D1-D5$ brane intersection since
parallel $D1$-$D1$ and $D5$-$D5$ branes satisfy the same
supersymmetry algebra as the single $D1$- and $D5$-brane which
had already been given in \ct{kim}. Therefore only DN and DD
coordinates are involved in this brane configuration.

The conserved kinematical supersymmetry is given by
\be \la{mode-kin-d+d+-sym}
q^+ = 2 \sqrt{2p^+} I_+ \widehat{S}_0
\ee
where
\be \la{s0++}
\widehat{S}_0= \sqrt{\frac{\sinh \pi \mu |\a|}{\pi \mu |\a|}}
e^{\half \pi \mu |\a| \Omega_0 \Pi} S_0.
\ee
And the dynamical supercharge takes the form
\be \la{d5-dyn-mode}
\sqrt{2p^+}q^{-} =  2I_+ \sum_{\k}\sum_{I=\{i^\prime, r^\prime \}}
\Bigl( c_\k  \a_{-\k}^I \gamma^I S_{\k}
- \frac{i \mu}{2c_\k \omega_\k}\a_{-\k}^I \Omega_0 \gamma^I \Pi S_{\k}
\Bigr),
\ee
where we assumed that the $D1, D5$-branes are placed at origin for
simplicity. Here the mode number
$\k$ should be understood as nonzero integers for $I=r^\prime$
and real numbers satisfying the second equation
in Eq. \eq{trans-eq} for $I=i^\prime$.

The supersymmetry algebra in this case is of the form:
\bea \la{d+d+-susy-alg}
&& \{q^+_a, q^+_b\}= 2p^+ (I_+)_{ab}, \\
&& \{ q^+_a, q^-_{\ad} \} = 0,\\
&& \{q^-_{\ad}, q^-_{\bd}\}= 2H (I_+)_{\ad\bd},
\eea
where the hamiltonian is given by
\be \la{d5-h-mode}
 2p^+ H= \sum_{n \neq 0}
\Bigl( \half \a_{-n}^{r^\prime}  \a_n^{r^\prime}
+ 2 \omega_n S_{-n} I_+ S_n \Bigr)
+ \sum_{\k} \Bigl( \half \a_{-\k}^{i^\prime} \a_\k^{i^\prime}
+ 2 \omega_\k S_{-\k} I_- S_\k \Bigr).
\ee

%%%%%%%%%%%%%%%%%%%%%%%%%%%%%%%%%%%%%%%%%%%%%%%%%%%%%%%%%%%%%%%%%%%%%%
\section{Discussion}
%%%%%%%%%%%%%%%%%%%%%%%%%%%%%%%%%%%%%%%%%%%%%%%%%%%%%%%%%%%%%%%%%%%%%%

In this paper we studied intersecting D-branes in the type IIB
plane wave background using Green-Schwarz superstring action. It
turned out that this method is quite elegant and sufficiently
powerful since the spacetime supersymmetry of intersecting
D-branes and stretched open strings between them are manifest
without any GSO projection. However we should confess that this
method also has a similar defect
appearing in the boundary state formalism \ct{gg}.
First of all, it is not obvious how to
treat intersecting D-branes with $\sharp_{ND} = 2,6$
which are the cases completely breaking supersymmetry. In
addition, $D3$-$D1$ intersection with $\sharp_{ND} = 4$ is invisible since the
light-like coordinate $X^-$ always has to
satisfy Neumann boundary condition in the light-cone gauge \ct{dabholkar,kim}.
Since we have used the light-cone gauge in the plane wave
background, $D3$-$D1$ intersection is missed in the Tables 1-2
although it is not obvious whether they preserve the
supersymmetry.

In this work we have not considered oblique D-branes which was
recently discussed in \ct{hikida,ggsns}. It was argued \ct{chu1} that the isometry
in the plane wave background \eq{pp-metric} is indeed $\so \times \sop \times Z_2$
where the $Z_2$ symmetry interchanges simultaneously the two
$SO(4)$ directions
\be \la{z2}
Z_2 : (x^1,x^2,x^3,x^4) \leftrightarrow (x^5,x^6,x^7,x^8).
\ee
The boundary condition of the oblique D-branes is invariant under
the $Z_2$ involution \eq{z2}. Thus it should be interesting to know whether
the oblique D-branes can be understood by appropriately utilizing
the $Z_2$ symmetry and to classify the complete list of the
oblique D-branes and their intersections.

The most interesting feature of intersecting D-branes is the
appearance of chiral fermions on the intersection of D-branes \ct{bdl}.
Since we have used spacetime fermions to study intersecting
D-branes, the appearance of the chiral fermions should be more
directly seen compared to the NSR formulation which relies on
vertex operator construction and GSO projection. An interesting
question is whether the chiral fermions can also appear on
intersecting D-branes in a plane wave background. We have seen
that fermion zero modes on $D_-$-branes become massive in the
plane wave background and even for the cases with massless fermion
zero modes, e.g., $D_+ - D_+$ intersections,
D-brane intersection to realize $\CN =1$ chiral multiplet seems to
be impossible, as illustrated in Table 2. Thus the chiral fermion
seems to be highly implausible at least in the type IIB plane wave
background.

It was known \ct{park,witten} that supersymmetric intersections with $\sharp_{ND} = 2, 6$
can be allowed when a suitable B-field is turned on. In addition
the BPS bound states of D-branes in the presence of Neveu-Schwarz
B-field is T-dual to
D-branes intersecting at angles \ct{bdl}. It will be interesting to study
how to generalize Green-Schwarz worldsheet formulation for
intersecting D-branes to the case of the presence of the
Neveu-Schwarz B-field. We hope to report our progress along this
direction elsewhere.

\section*{Acknowledgments}

We are supported by the grant from the Interdisciplinary Research
Program of the KOSEF (No. R01-1999-00018) and a special grant of
Sogang University. BHL is supported by the Korean Research
Foundation Grant KRF D00027. HSY was supported by the Brain Korea
21 Project in 2003. BHL thanks ITP at University of Hannover and
KIAS for the hospitality and O. Lechtenfeld for helpful
discussions.

\appendix

\section*{Appendix}

%%%%%%%%%%%%%%%%%%%%%%%%%%%%%%%%%%%%%%%%%%%%%%%%%%%%%%%%%%%%%%%%%%%%%%%
\section{Supersymmetry Algebra of $D_-3 - D_-5$ Intersection}
%%%%%%%%%%%%%%%%%%%%%%%%%%%%%%%%%%%%%%%%%%%%%%%%%%%%%%%%%%%%%%%%%%%%%%%

To illustrate the superalgebra \eq{dyn-qq} explicitly,
let us consider a $D_-3$-brane oriented along $X^{1,2}$ directions
and intersecting with a $D_-5$-brane oriented along $X^{2,3,4,5}$.
In this case, we have
\be \la{pi-omega}
\Pi \Omega_0= - \gamma^{34}, \qquad \Omega_0^T \Omega_\pi = -
\gamma^{1345}.
\ee
 We will use the indices $i^\prime, j^\prime = 3,4,5$ for the DN coordinates
and the indices $r^\prime, s^\prime = 6,7,8$ for the DD coordinates. The following
Fierz identity is useful to compute the
supersymmetry algebra \eq{dyn-qq}:
\be \la{fierz}
S_1^{\pm a} S_2^{\pm b} = \frac{1}{8} \delta_{ab} S_1^\pm S_2^\pm +
\frac{1}{16}\gamma^{IJ}_{ab} S^\pm_1\gamma^{IJ}S^\pm_2
+\frac{1}{384} \gamma^{IJKL}_{ab} S^\pm_1 \gamma^{IJKL} S^\pm_2,
\ee
where $S_A^{\pm a} = (I_\pm S^A)^a$ and $S^A$ are the spinors with positive
chirality.

For the brane configuration in Eq. \eq{pi-omega}, the algebra of dynamical
supersymmetry is given by
\be \la{comm-qq}
\{ q^-_{\ad}, q^-_{\bd} \} = \frac{2}{p^+} \{ \sqrt{2p^+} Q_{\ad}^{-2},
\sqrt{2p^+} Q_{\bd}^{-2} \}
\ee
where
\bea \la{Q2}
\sqrt{2p^+} Q_{\ad}^{-2} &=& ( p_0^2 \gamma^2 \Omega_0^T
+ \mu x_0^2 \gamma^2 \Pi) I_- S_0 \\
&& + \sum_{n \neq 0} \Bigl( c_n ( \a_{-n}^2 \gamma^2 + \a_{-n}^{r^\prime} \gamma^{r^\prime})
I_+ S_{n} - \frac{i \mu}{2c_n \omega_n}
( \a_{-n}^2 \gamma^2 - \a_{-n}^{r^\prime} \gamma^{r^\prime}) \Pi \Omega_0 I_+ S_{n} \Bigr) \xx
&& + \sum_{\k} \Bigl( c_\k ( \a_{-\k}^1 \gamma^1 + \a_{-\k}^{i^\prime} \gamma^{i^\prime})
I_- S_{\k} - \frac{i \mu}{2c_\k \omega_\k}
( \a_{-\k}^1 \gamma^1 - \a_{-\k}^{i^\prime} \gamma^{i^\prime})
\Pi  \Omega_0 I_- S_{\k} \Bigr). \nonumber
\eea
It is straightforward to calculate the anticommuatator
\eq{comm-qq} for the supercharge \eq{Q2} using the rule
\be \la{comm-rule}
\{ B_1 F_1, B_2 F_2 \} = [B_1, B_2] F_1 F_2
+ B_2 B_1 \{ F_1, F_2 \}
\ee
for bosonic modes $B_1, B_2$ and fermionic modes $F_1, F_2$. The
overall calculation is similar to the single D-brane case \ct{kim}.
However, one should pay attention to the fact that the spinors
$S_\nu^\pm \;(\nu=n, \k)$
are now used instead of $S_\nu$.

Let us explain the useful relations used in the calculation
of Eq. \eq{comm-qq}:
\bea \la{formula1}
&& f^{IJ} I_\pm \gamma^{IJ} I_\pm = (2 f^{1 i^\prime} \gamma^{1 i^\prime}
+ 2 f^{2 r^\prime} \gamma^{2 r^\prime}
+  f^{i^\prime j^\prime} \gamma^{i^\prime j^\prime}
+  f^{r^\prime s^\prime} \gamma^{r^\prime s^\prime} ) I_\pm , \\
\la{fomula2}
&& \frac{1}{48} \sum_{\nu}  f^{IJKL}_{\ad \bd} S^\pm_{-\nu} \gamma^{IJKL} S^\pm_\nu
= \mp  \sum_\nu f^{1345}_{\ad \bd} S^\pm_{-\nu} S^\pm_\nu, \\
\la{fomula3}
&& (\gamma^{15} \Pi\Omega_0 I_+)_{\ad \bd} S^\pm_{-\nu} \gamma^{15}
S^\pm_\nu = \pm (\gamma^{34} \Pi\Omega_0 I_+)_{\ad \bd} S^\pm_{-\nu} \gamma^{34}
S^\pm_\nu, \\
\la{fomula4}
&& (\gamma^{2 r^\prime} \Pi\Omega_0 I_+)_{\ad \bd} S^\pm_{-\nu} \gamma^{2 r^\prime}
S^\pm_\nu = \mp \half (\gamma^{r^\prime s^\prime} \Pi\Omega_0 I_+)_{\ad \bd}
S^\pm_{-\nu} \gamma^{r^\prime s^\prime}
S^\pm_\nu,
\eea
where $f^{IJ}$ is a fermion bilinear and $f^{IJKL}_{\ad \bd}$ is a
product of gamma matrices. In Eqs. \eq{fomula2} and \eq{fomula4}, $\gamma_{\ad
\bd}= - \delta_{\ad \bd}$ was used according to Eq. \eq{gamma9}.
In addition, one needs to use the following zeta function
regularization to get the algebra \eq{dyn-qq} correctly:
\be \la{zeta}
\sum_{n \in \bfz} 1 = \sum_{\k \in \bfz + \half} 1 =0.
\ee
After carefully using these facts, one can finally get the
supersymmetry algebra \eq{dyn-qq} with the rotation generators $J^{34}_{I} \in SO(2)$
and $J^{r^\prime s^\prime}_{II} \in SO(3)$.

\newpage

%%%%%%%%%%%%%%%%% Journal Macros %%%%%%%%%%%%%%%%%%%%%%%%%%%

\nc{\np}[3]{Nucl. Phys. {\bf B#1}, #2 (#3)}

\nc{\pl}[3]{Phys. Lett. {\bf B#1}, #2 (#3)}

\nc{\prl}[3]{Phys. Rev. Lett. {\bf #1}, #2 (#3)}

\nc{\prd}[3]{Phys. Rev. {\bf D#1}, #2 (#3)}

\nc{\ap}[3]{Ann. Phys. {\bf #1}, #2 (#3)}

\nc{\prep}[3]{Phys. Rep. {\bf #1}, #2 (#3)}

\nc{\ptp}[3]{Prog. Theor. Phys. {\bf #1}, #2 (#3)}

\nc{\rmp}[3]{Rev. Mod. Phys. {\bf #1}, #2 (#3)}

\nc{\cmp}[3]{Comm. Math. Phys. {\bf #1}, #2 (#3)}

\nc{\mpl}[3]{Mod. Phys. Lett. {\bf #1}, #2 (#3)}

\nc{\cqg}[3]{Class. Quant. Grav. {\bf #1}, #2 (#3)}

\nc{\jhep}[3]{J. High Energy Phys. {\bf #1}, #2 (#3)}

\nc{\hep}[1]{{\tt hep-th/{#1}}}

%%%%%%%%%%%%%%%%%%%%%%%%%%%%%%%%%%%%%%%%%%%%%%%%%%%%%%%%%%%

\end{document}